\documentclass[twocolumn, pre, showpacs, english, preprintnumbers, amsmath, amssymb, superscriptaddress, aps,longbibliography]{revtex4-2}
\usepackage[utf8]{inputenc}
\usepackage{graphicx}
\usepackage[dvipsnames]{xcolor}
\usepackage{appendix}
\newcommand{\orcid}[1]{\href{https://orcid.org/#1}{\includegraphics[width=8pt]{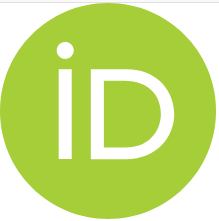}}}

\renewcommand\vec{\boldsymbol}

\usepackage[colorlinks,urlcolor=goodgreen,citecolor=blue,linkcolor=goodred]{hyperref}

\definecolor{orange}{rgb}{1,0.5,0}
\definecolor{goodgreen}{rgb}{0.1,0.5,0}
\definecolor{goodred}{rgb}{0.7,0,0}
\let\oldepsilon\epsilon \let\epsilon\varepsilon \let\varepsilon\oldepsilon
\makeatother
\begin{document}
\title{Spectroscopic signature of spin triplet odd-valley  superconductivity in two-dimensional materials}
\author{T.H. Kokkeler\orcid{0000-0001-8681-3376}}
\email{tim.kokkeler@dipc.org}
\affiliation{Donostia International Physics Center (DIPC), 20018 Donostia--San Sebasti\'an, Spain}
\affiliation{University of Twente, 7522 NB Enschede, The Netherlands}

\author{Chunli Huang\orcid{0000-0002-0928-6266}}
\email{chunli.huang@uky.edu}
\affiliation{Department of Physics and Astronomy, University of Kentucky, Lexington, Kentucky 40506-0055, USA}

\author{F.S. Bergeret\orcid{0000-0001-6007-4878}}
\email{fs.bergeret@csic.es}
\affiliation{Centro de F\'isica de Materiales (CFM-MPC) Centro Mixto CSIC-UPV/EHU, E-20018 Donostia-San Sebasti\'an,  Spain}
\affiliation{Donostia International Physics Center (DIPC), 20018 Donostia--San Sebasti\'an, Spain}

\author{I. V. Tokatly\orcid{0000-0001-6288-0689}}
\email{ilya.tokatly@ehu.es}
\affiliation{IKERBASQUE, Basque Foundation for Science, 48009 Bilbao, Basque Country, Spain}
\affiliation{Donostia International Physics Center (DIPC), 20018 Donostia--San Sebasti\'an, Spain}
\affiliation{Nano-Bio Spectroscopy Group and European Theoretical Spectroscopy Facility (ETSF), Departamento de Polímeros 
y Materiales Avanzados: Física, Química y Tecnología, Universidad del País Vasco, 20018 Donostia-San Sebastián, 
Basque Country, Spain}

\begin{abstract}
Motivated by recent discoveries of superconductivity in lightly-doped multilayer graphene systems, we present a low-energy model to study superconductivity in 2D  materials whose Fermi surface consists of two valleys  at $\pm \vec{K}$-points. We assume a triplet odd-valley superconducting order with a pair potential that is isotropic in each valley but has a different  sign in the two different valleys.
Our theory predicts the emergence of an almost flat band of edge states centered at zero energy for certain edge orientations. As a result, a prominent experimental signature of this type of superconductivity is the presence of a large zero-energy peak in the local density of states near specific edges. 
 The results of the effective low-energy theory are confirmed by numerically analyzing a specific microscopic tight-binding realization of odd-valley superconductivity, f-wave superconductivity on a honeycomb lattice in a ribbon geometry. Our work provides a test for odd-valley superconductivity through edge spectroscopy.

\end{abstract}

\maketitle
\textit{Introduction:--}
Superconductivity emerges in multilayer graphene stacks that are perturbed by a magic-angle twist potential \cite{cao2018unconventional, lu2019superconductors,stepanov2020untying,szabo2022metals,yankowitz2019tuning,lian2019twisted,park2022robust,hao2021electric,zhang2022promotion,chen2019signatures,park2021tunable}  or by a strong electric displacement field \cite{zhou2021superconductivity, zhou2022isospin,holleis2023ising,zhang2022spin,heikkila2022surprising}. These systems have received a lot of attention lately. However, it is still unclear if and how the superconducting order parameter $\Delta$ changes sign in their hexagonal Brillouin zones. The momentum-dependence of the pair potential usually reflects the underlying pairing mechanism and thus can be used to constrain microscopic theories of superconductivity \cite{scalapino2012common,crepel2022unconventional,CeaSC1,CeaSC2,pantaleon2022superconductivity,Ghazaryan2021,jimeno2022superconductivity,pantaleon2023superconductivity,sainz2022junctions,guinea2012odd}. In this letter, we identify a smoking-gun local tunneling spectroscopy signature to differentiate a spin-triplet odd-valley superconductor from a spin-singlet even-valley superconductor and illustrate it with a simple tight-binding model calculation.

\begin{figure}
    \centering
    \includegraphics[width = 8.6cm]{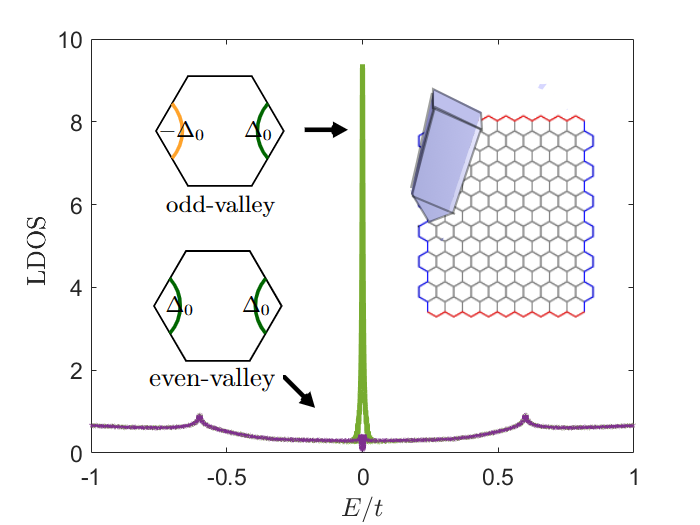}
    \caption{Spin triplet odd-valley superconductivity in graphene leads to a large peak in the local density of states near armchair edges. The result is calculated using a tight-binding model presented later in the paper with superconducting-gap/Fermi energy $\Delta/\mu = 0.05$.
    Features at $\Delta/\mu\approx 0.6$ are 
    not related to superconductivity but  to the underlying band structure. The smearing parameter in the calculation of the density of states was set to $\delta/\Delta = 0.1$. 
     }
    \label{fig:MainFigure}
\end{figure}

The Fermi surface of lightly-doped multilayer graphene systems is centered at the two valleys $(K,K')$, i.e.~the two inequivalent corners of the hexagonal Billouin zone.
When $\Delta$ changes sign in a single valley and has nodal points on the Fermi surface,
the density-of-states (DOS) decreases continuously to zero as the energy approaches the Fermi level.
Such DOS profile can be probed by scanning-tunneling spectroscopy and leads to a $V-$shaped tunneling spectrum \cite{kashiwaya2000tunnelling,kim2022evidence}. By contrast,
if there are no nodes on the Fermi surface, it leads to the usual $U$-shaped local tunneling spectrum irrespective of the relative sign of $\Delta$ in the two valleys because in either case the excitation energy of all Bogoliubov quasiparticles is gapped. However, as we demonstrate below, the superconducting gap of an odd-valley superconductor can close at the boundary of the two-dimensional materials while the spectrum of an even-valley superconductor remains gapped. Moreover, the dispersion of the edge states in an odd-valley superconductor is anomalously flat, with $\omega\sim\Delta^2/\mu$ where $\mu$ is the chemical potential and these in gap states lead to a large local DOS as shown in Fig.~\ref{fig:MainFigure}. On the contrary, for even-valley superconductors there are no edge states and a zero energy peak is absent, see Fig.~\ref{fig:MainFigure}. Since the LDOS can be probed using STM measurements, our calculations provide a tool for identifying odd-valley superconductivity.

In what follows, we calculate the edge state dispersion of a  superconductor with spin-triplet odd-valley pairing, under the assumptions that $\Delta\ll\mu$ and that the Fermi surface consists of a single band. Using a universal property of Fermi liquids -- the excitation energy is particle-hole symmetric at the Fermi surface -- we demonstrate that the edge spectrum is macroscopically concentrated around zero energy.
In the second part of the Letter, 
 we use a tight-binding Hamiltonian to calculate the edge state dispersion and to confirm the results of our low-energy theory. 

\textit{Low energy model:-- }
We start with a general 2D multiband superconductor described by the Hamiltonian 
\begin{align}\label{eq:LowEnergyHamiltonian}
    H = \sum_{\alpha}\begin{bmatrix}
\xi_{\alpha,\vec{k}}&\Delta_{\alpha,\vec{k}}\\\Delta_{\alpha,\vec{k}}&-\xi_{\alpha,-\vec{k}}
    \end{bmatrix}\otimes|u_{\alpha,\vec{k}}\rangle\langle u_{\alpha,\vec{k}}|\;\;.
\end{align}
Here $\xi_{\alpha,\vec{k}}, \Delta_{\alpha,\vec{k}}$ are the quasiparticle energy and pair potential and $\alpha,\vec{k}$ are the band index and crystal momentum respectively, $|u_{\alpha,\vec{k}}\rangle\langle u_{\alpha,\vec{k}}|$ is the band projector. The corresponding Green function $G_{\vec{k}}(\omega)=(\omega - H)^{-1}$ in momentum space reads
\begin{align}\label{eq:LowEnergyGreensfunction}
    G_{\vec{k}}(\omega) 
    = \sum_{\alpha}\frac{(\omega\mathbf{1}_{\tau}-\xi_{\alpha,\vec{k}}\tau_{3}+\Delta_{\alpha,\vec{k}}\tau_{1})}{\omega^{2}-\xi_{\alpha,\vec{k}}^{2}-\Delta_{\alpha,\vec{k}}^{2}}\otimes |u_{\alpha, \vec{k}}\rangle\langle u_{\alpha,\vec{k}}|,
\end{align}
where $\tau_{i}\, (i = 1,2,3)$, and $\mathbf{1}_{\tau}$ are, respectively, the Pauli and the identity matrices spanning the Nambu space. We assume that the Fermi surface crosses only one of the bands, and that the energy difference between bands is much larger than the pairing energy, such that interband correlations can be neglected.

We are interested in describing the bound states  at  the  edge of an  odd-valley superconductor, that is, a superconductor whose pair potential has a different sign in each valley.
The sharp edge, located at the line $x=0$, is modelled by adding to the periodic potential of the crystal a 1D delta-potential, $U(\mathbf{r})= V\delta(x)\tau_{3}$, and taking the limit $V\xrightarrow{}\infty$, which automatically imposes the wave functions to vanish at $x = 0$, effectively making it an edge.

Wave functions $|\psi(x)\rangle$ of the edge states are obtained using the Lippmann-Schwinger equation \cite{lippmann1950variational},
\begin{align}
\label{eq:LS_wf}
    |\psi(x)\rangle = V\tau_{3}G_{x,k_{y}}(\omega)|\psi(x = 0)\rangle.
\end{align}
Here $G_{x ,k_{y}}$ is the Fourier transform, with respect to $k_x\xrightarrow{}x$, of the  Green's function in Eq. (\ref{eq:LowEnergyGreensfunction}):   
\begin{align}
    G_{x,k_{y}}(\omega) &= \int_{BZ} e^{ik_{x}x}G_{k_{x},k_{y}}(\omega) \frac{dk_x}{2\pi} 
    \label{eq:Gx}
\end{align}

\begin{figure}
    \centering
    \includegraphics[width = 8.6cm]{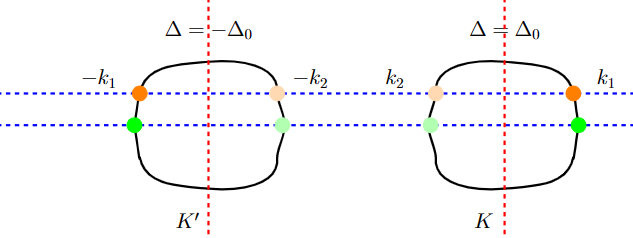}
    \caption{Schematic of a Fermi surface with two disconnected components and an odd-valley pair potential. The dashed lines correspond to lines of integration      for the Fourier transforms. The orientation of the edge determines which line of integration should be taken. The blue lines cross both valleys and thus edges with this orientation will have edge states. The red lines cross only one valley, and therefore edges with this orientation do not exhibit edge states.    }\label{fig:GrapheneP}
\end{figure}
From Eq. (\ref{eq:LS_wf}), the energy of the  bound states is determined  by the equation
\begin{align}\label{eq:LSxky}
    \text{det}\left( \, V^{-1}-\tau_{3}G_{x=0,k_{y}}(\omega)   \right)= 0.
\end{align}
Equations (\ref{eq:LS_wf}) and (\ref{eq:LSxky}) depend parametrically on $k_y$. This dependence determines the dispersion of the edge states. In order to solve this equation and then  construct the wave function of the edge states,  one needs to evaluate the integral 
in Eq. (\ref{eq:Gx}). The integration runs  over $k_x$ at fixed $k_y$, {\it i.~e.} over straight lines in $k$-space.  By assumption, the Fermi surface consists of two disconnected pockets (valleys) surrounding two distinct points in the Brillouin zone $\vec{K}$ and $-\vec{K}$, related by the inversion operation, as typically occurs in graphene-related materials, see Fig. \ref{fig:GrapheneP}.
The shape of the Fermi surfaces around these two points is arbitrary. We assume  that the pair potential is  constant within a single valley, but changes sign between the valleys.
As mentioned above,  due to the Pauli exclusion principle, this  odd-valley superconductivity corresponds to a triplet state. \footnote{The two valleys are centered at $\pm K$-points in the Brillouin zone so that the triplet odd-valley pair potential is odd-parity, that is $\Delta(-(\vec{K}+\delta\vec{k})) = -\Delta(\vec{K}+\delta\vec{k})$.  This distinguishes this type of superconductivity from multi-orbital superconductivity with $s_{\pm}$-pairing \cite{onari2099surface,burmistrova2015josephson,fukaya2018interorbital,burmistrova2013theory,sato2011topology}, which is predicted to exist in pnictides \cite{golubov2009andreev,mashkoori2019impact}.}.

When the pair potential is much smaller than both the Fermi energy, $\Delta\ll\mu$, \cite{zhou2021superconductivity,zhou2022isospin} and the energy separation from other bands, the integral in Eq.~(\ref{eq:Gx}) is dominated by the band crossing the Fermi level and all remote bands may be ignored. We therefore leave only this relevant band in the Green function of Eq.~(\ref{eq:LowEnergyGreensfunction}), and everywhere below drop the band index by writing $|u_{\vec{k}}\rangle, \xi_{\vec{k}}$ and $\Delta_{\vec{k}}$.

The result of the integration depends on the orientation of the integration  lines in $k$-space, which, in turn, are determined by the normal to the edge as by construction they are orthogonal to $x$-axis.  We focus here on two edge orientations, indicated by the red and blue lines in Fig. \ref{fig:GrapheneP}. In the case of graphene they correspond to zig-zag and armchair edges, respectively, see Fig.~\ref{fig:MainFigure}. 
In the red case, the  lines  of integration  may cross the Fermi contour only in a single  valley.   Because the pair potential is assumed to be isotropic across the valley, this situation is equivalent to the edge of a conventional s-wave superconductor, and therefore the system does not  exhibit edge states at such edges.

More interesting is the orientation of the edge corresponding to the blue lines in Fig.  \ref{fig:GrapheneP}. 
In this case, the line of integration may cross either none, or both valleys. 
In the former case,  one can verify from the Lippmann-Schwinger equation that  no  edge states appear. Therefore, we focus on values of $k_{y}$ for which the line of integration crosses both valleys. 

To compute the integral over $k_x$ we notice that in the case $\Delta\ll\mu$ 
the main contribution is from  momenta close to the Fermi momentum.
Therefore, under the standard assumptions of Fermi liquid theory, we linearize the spectrum around the points $\vec k_n$ where the line of integration crosses the Fermi contour. As a result, the integral in Eq.~(\ref{eq:Gx}) is transformed to the sum of the integrals over $\xi_{\vec k_n}$ at each crossing point. 
Specifically, $G_{x = 0,k_{y}}(\omega)$ is given by
\begin{align}
  \label{eq:Gqc}
   G_{x=0,k_{y}}\approx
   -\sum_{n=1,2}\frac{1}{2v_{n}}
    \frac{\omega}{\sqrt{\Delta^{2}-\omega^{2}}}\mathbf{1}_{\tau}\otimes|u_{\vec{k}_{n}}\rangle\langle u_{\vec{k}_{n}}|\; , 
\end{align}
where $v_{1,2}=|\partial\xi_{\vec k}/\partial k_x|_{\vec k_{1,2}}$ are $x$-components of the Fermi velocities at the points $\vec k_{1,2}$ at which the line of integration crosses the Fermi contour in the K-valley, see Fig.~\ref{fig:GrapheneP}. 
Thus,  at $x = 0$ the Green's function is proportional to the unit matrix in Nambu space and comes from the term $\sim\omega\mathbf{1}_\tau$ in Eq.~(\ref{eq:LowEnergyGreensfunction}). The term $\sim\xi_{\vec k}\tau_3$ vanishes upon $\xi$-integration due to the particle-hole symmetry inherent to the linearized spectrum. Finally, the $\Delta_{\vec k}\tau_{1}$ contribution vanishes after summation over the valleys due to the valley-antisymmetry of the pairing potential.

By substituting Eq.~(\ref{eq:Gqc}) into Eq.~(\ref{eq:LS_wf}) at $x=0$, and projecting it onto the Bloch states $|u_{\vec k_1}\rangle$ and $|u_{\vec k_2}\rangle$ we get a 4$\times$4 problem for two Nambu spinors $\langle u_{\vec k_{1,2}}|\psi(0)\rangle$. By evaluating the determinant of the corresponding 4$\times$4 matrix in Eq.~(\ref{eq:LSxky}) and taking the limit $V\to\infty$, it follows that there exist four edge states with $\omega(k_y) = 0$,  for all $k_y$. Importantly, the zero energy of the edge states and the absence of dispersion,  
while true with very high accuracy, is nonetheless an approximate property. It is a consequence of the approximate electron-hole symmetry that is controlled by the parameter $\frac{\Delta}{\mu}\ll 1$.

To find the wave functions of the edge states we compute $G_{x,k_{y}}(\omega)$ using the same approximations, and insert the result into Eq.~(\ref{eq:LS_wf}). This yields four states which in the limit $V\to\infty$ naturally split into two pairs of states localized on the opposite sides of the barrier.
The wave functions of these four edge states with energies $ \omega(k_y)=0+O(\Delta^2/\mu)$ read
\begin{align}
    \Psi_{L1,2} &= \Bigg(\sin k_{1,2}x\begin{bmatrix}
     1 \\\pm i
    \end{bmatrix}\otimes |u_{k_{1,2}}\rangle e^{\kappa_{1,2} x}+\nonumber\\&\sin k_{2,1}x\langle u_{k_{2,1}}|u_{k_{1,2}}\rangle \begin{bmatrix}
        1\\\mp i
    \end{bmatrix}\otimes |u_{k_{2,1}}\rangle e^{\kappa_{2,1}x}\Bigg)\Theta(-x),\\
    \Psi_{R1,2} &= \Bigg(\sin k_{1,2}x\begin{bmatrix}
     1\\\mp i
    \end{bmatrix}\otimes |u_{k_{1,2}}\rangle e^{-\kappa_{1,2} x}+\nonumber\\&\sin k_{2,1}x \langle u_{k_{2,1}}|u_{k_{1,2}}\rangle\begin{bmatrix}
        1\\\pm i
    \end{bmatrix}\otimes |u_{k_{2,1}}\rangle e^{-\kappa_{2,1} x}\Bigg)\Theta(x)\;, 
\end{align}
where the upper sign in $\pm,\mp$ corresponds to $\Psi_{L,R1}$ and the lower sign to $\Psi_{L,R2}$, and where $\kappa_n=\sqrt{\Delta^2-\omega^2}/v_n$.
The states $\Psi_{L1,2}$ and   $\Psi_{R1,2}$ are localized on the left- and right-side of the potential wall respectively, as indicated by the Heaviside functions $\Theta(\mp x)$. Therefore, to the leading order in $\Delta/\mu\ll 1$, each physical edge supports two degenerate zero energy edge states for each $k_{y}$. Lifting the degeneracy and the appearance of a weak dispersion as a higher order effect in $\Delta/\mu$ is analyzed in detail in the supplemental material \cite{supplemental}, and also discussed below for a specific lattice model of odd-valley superconductor.

The above results  can be directly applied to graphene or other materials with hexagonal lattices. 
 Armchair edges with  $\Delta \vec{K} = 2K_{x}\hat{x}$, correspond to the blue lines of integration in Fig. \ref{fig:GrapheneP}. In this case, our theory  predicts the existence of a flat band of zero-energy edge states. This will manifest as a large peak in the density of states, localized at the edge over a coherence length. 
 On the other hand, zigzag edges with $\Delta \vec{K} = 2K_{y}\hat{y}$,  correspond to the red lines of integration in Fig. \ref{fig:GrapheneP}. In this case, no edge states are expected. We emphasize that this difference between armchair and zigzag edges is due to the normal of the surface being parallel or perpendicular to $K-K'$ line, not to the exact shape of the edges.
 These two features can be used to  unequivocally characterize the odd-valley superconductivity in graphene-like materials.  The superconductors described by our low-energy model are topologically trivial, since though they are odd-parity, the Fermi surface does not enclose time-reversal invariant momenta \cite{sato2011topology}. If additionally mirror-symmetry is present, the superconductors may have a mirror topology, such as in odd-layer graphene stacks \cite{phong2023mirror}.

\begin{figure*}[!t]
    \centering
    {\hspace*{-2em}\includegraphics[width = 5.6cm]{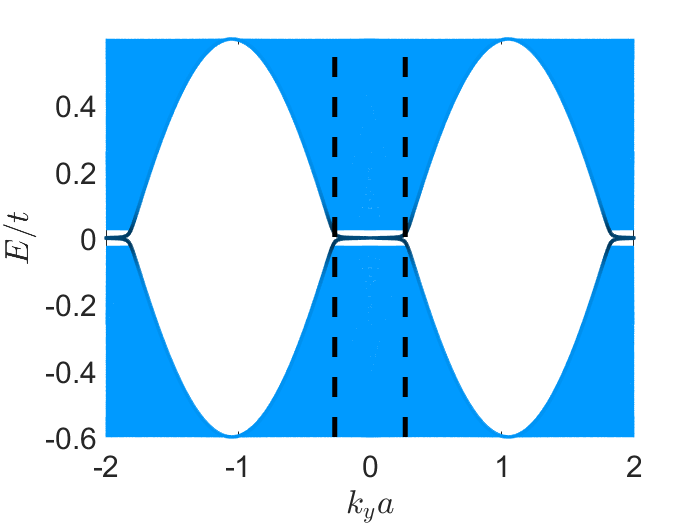}}
    \hfill
    {\hspace*{-2em}\includegraphics[width = 5.6cm]{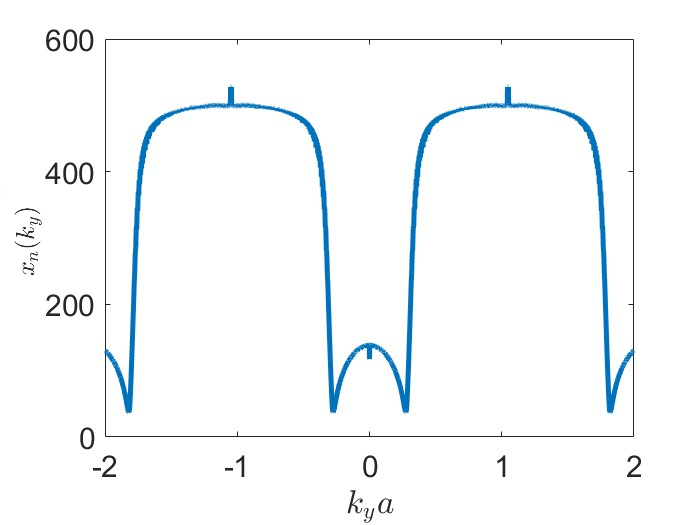}}
    \hfill
    {\hspace*{-2em}\includegraphics[width = 5.6cm]{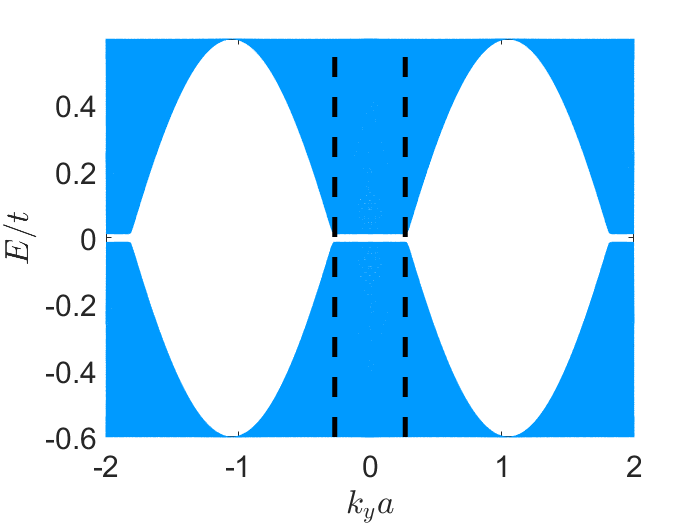}}
    \hfill
    {\hspace*{-1.5em}\includegraphics[width = 5.6cm]{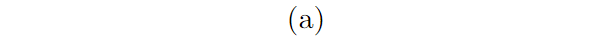}}
    \hfill
    {\hspace*{-2em}\includegraphics[width = 5.6cm]{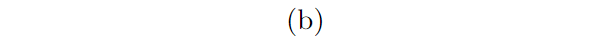}}
    \hfill
    {\hspace*{-2em}\includegraphics[width = 5.6cm]{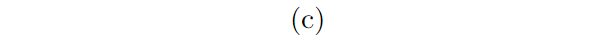}}
    \caption{\textit{a:} 
   Dispersion for odd-valley superconductivity calculated using the tight-binding model for $\Delta/\mu = 0.05$ and  $\mu/t = 0.4$, plotted over one full Brillouin zone. The edge states are clearly separated in energy from the other states. They are flat near $k_{y} = 0$ and merge with the band near $k_{y} = k_{F}$, indicated by the dashed lines.
   \textit{b:} The average value of the position operator for the edge states as a function of $k_y$.  The states are well localized for $k_{y}\ll k_{F}$, indicated by the dashed lines in panel (a). The localization length is smallest near the Fermi surface $k_{y} = k_{F}$ and  the states become delocalized as their energy approaches the bulk gap for $k_{y}>k_{F}$. Here $a$ is the carbon-carbon distance.
    \textit{c:}  Dispersion for even-valley superconductivity calculated using the tight-binding model for $\Delta/\mu = 0.05$ and $\mu/t = 0.4$. The bulk dispersion is similar to odd-valley superconductivity, but edge states are absent.
    \label{fig:DispersionTB}}    
\end{figure*}

\textit{Lattice model--}
As a microscopic illustration of the above low-energy theory, we use a specific lattice realization of an odd-valley superconductor to demonstrate the appearance of the massively degenerate edge states.
We focus on a honeycomb tight-binding lattice and consider a ribbon with infinite extension in the $y$-direction and restricted by two armchair edges in the $x$-direction. 
In order to generate an effective odd-valley superconducting order parameter, we consider pairing only in the next-nearest ($AA$) sublattice. This type of intra-sublattice pairing potential has been microscopically studied in Refs.~\cite{crepel2022unconventional,huang2022pseudospin}.

We consider spinless electrons. The Hamiltonian for a given Bloch momentum $k_{y}$ reads:
\begin{align}
    H(k_{y}) &=  \sum_{i}-\mu (a_{i,k_{y}}^{\dagger}a_{i,k_{y}}+b_{i,k_{y}}^{\dagger}b_{i,k_{y}})+\nonumber\\&t(e^{-ik_{y}}a_{i,k_{y}}^{\dagger}b_{i,k_{y}}+e^{i\frac{1}{2}k_{y}}a_{i,k_{y}}^{\dagger}b_{i+1,k_{y}}+e^{i\frac{1}{2}k_{y}}a_{i,k_{y}}^{\dagger}b_{i-1,k_{y}})\nonumber\\&+\Delta\Bigg((a_{i-2,k_{y}}^{\dagger}a^{\dagger}_{i,k_{y}}-a_{i+2,k_{y}}^{\dagger}a^{\dagger}_{i,k_{y}})\nonumber\\& +(-2\cos\frac{3}{2}k_{y})(a_{i-1,k_{y}}^{\dagger}a^{\dagger}_{i,k_{y}}-a_{i+1,k_{y}}^{\dagger}a^{\dagger}_{i,k_{y}})\nonumber\\&+(b_{i-2,k_{y}}^{\dagger}b^{\dagger}_{i,k_{y}}-b_{i+2,k_{y}}^{\dagger}b^{\dagger}_{i,k_{y}})+\nonumber\\&(-2\cos\frac{3}{2}k_{y})(b_{i-1,k_{y}}^{\dagger}b^{\dagger}_{i,k_{y}}-b_{i+1,k_{y}}^{\dagger}b^{\dagger}_{i,k_{y}})\Bigg)+\text{h.c.},
    \label{eq:TBH}
\end{align}
where the summation index $i$ runs over all unit cells from $i = 1$ to $i = 1024$, $a_{i}$ and $b_{i}$ are annihilation operators in unit cell $i$ on sublattices $A$ and $B$ respectively, $\mu$ is the chemical potential, $\Delta_{0}$ is the pair potential and, $\delta_{1} = a(0,-1),\delta_{2} = \frac{a}{2}(\sqrt{3},1),\delta_{3} = \frac{a}{2}(-\sqrt{3},1)$  are  vectors between nearest neighbours, while
$\chi_{1} = \delta_{3}-\delta_{2},\chi_{2} = \delta_{1}-\delta_{3},\chi_{3} = \delta_{2}-\delta_{1}$ are vectors between next-nearest neighbours. This model leads to f-wave superconductivity, studied in Refs. \cite{goudarzi2012tunneling,chou2021acoustic,chou2022acoustic,chou2021correlation,pangburn2022superconductivityI,pangburn2022superconductivityIII}. As shown in the supplemental material\cite{supplemental},
the tight-binding Hamiltonian, Eq.~\ref{eq:TBH} reduces to the low-energy model in Eq.~\ref{eq:LowEnergyHamiltonian} in the limit $t\gg\mu\gg\Delta$. For this model the band projectors $|u_{\vec{k}}\rangle\langle u_{\vec{k}}|$  are given by  $\frac{1}{2}(\mathbf{1}+\frac{k_{x}}{\sqrt{k_{x}^{2}+k_{y}^{2}}}\rho_{x}+\frac{k_{y}}{\sqrt{k_{x}^{2}+k_{y}^{2}}}\rho_{y})$, where  $\rho_{x,y}$ are  the first and second Pauli matrix in sublattice space.

Fig.~\ref{fig:DispersionTB}(a) shows the energy dispersion $E_n(k_y)$ v.s.~$k_y$  of our Hamiltonian, Eq.~\eqref{eq:TBH}.
The bulk spectrum (blue region) is  gapped, with a gap of the order of $\Delta$, and the edge states form an almost flat band around zero energy in the gap. As indicated using the dashed lines in \ref{fig:DispersionTB}(a), the flat band is well separated from the bulk for $k_{y}<k_{F}$ and merges with the bulk bands for $k_{y}\gg k_{F}$. The small
dispersion of the edge states for $k_{y} \ll k_{F}$ in Fig. \ref{fig:DispersionTB}(a) arises from particle-hole asymmetry which is controlled by the small parameter $\frac{\Delta}{\mu}$. This parameter is truly small for superconductors observed reported for Bernal bilayer graphene and rhombohedral trilayer graphene in Refs. \cite{zhou2021superconductivity,zhou2022isospin} where $T_c/T_F\sim10^{-3}$. We set $\frac{\Delta}{\mu}=0.05$ and $\frac{\mu}{t} = 0.4$ in our numerical calculations.
The result of this almost flat band is a strongly enhanced local density-of-states close to zero-energy, as shown in Fig~\ref{fig:MainFigure}.

Next, we compute the expectation value of the position operator for the eigenstate $\psi_{k_y,n}$:
\begin{align}
     x_n(k_y) = \frac{\int x|\psi_{k_y,n}|^{2}dx}{a\int |\psi_{k_y,n}|^{2}dx}.
\end{align} 
where $a$ is the carbon-carbon distance in our honeycomb lattice. Fig.~\ref{fig:DispersionTB}(b) shows $x_0(k_y)$ v.s.~$k_y$ where $n=0$ labels the positive energy states inside the superconducting gap.
 $x_0(k_y)$ is inversely correlated to the energy difference between the edge states and the bulk-state continuum. 
For $k_{y}\ll k_{F}$, $x_0(k_y)$ is very small compared to the width of the ribbon. In fact, $x_0(k_y=0) \sim t/\Delta$ and $x_0(k_y=k_F) \sim t(\Delta\mu)^{-1/2}$,  see the supplemental material \cite{supplemental}. In Fig. \ref{fig:DispersionTB}(c) we show that in-gap states are absent in the case of even-valley superconductivity, while the bulk spectrum is similar. This leads to the absence of a zero energy peak in the density of states as shown in Fig. \ref{fig:MainFigure}, confirming that the presence of a zero-energy peak signals unconventional superconductivity.

Since the ribbon Hamiltonian we consider is invariant under the mirror-plane $H(k_y)=H(-k_y)$, the counter-propagating edge states are located on the same position: $x_0(k_y)=x_0(-k_y)$. We found a generic disorder-induced transition matrix elements between them are finite $\langle \psi_{0,k_{y}}|\tau_3V|\psi_{0,-k_{y}}\rangle \neq0$ so the counter-propagating edge states will in fact be affected by edge imperfections. We may distinguish between smooth and sharp disorder. The states are robust against smooth disorder, since this does not mix the opposite valleys and thus leaves our conclusions unaltered. Sharp disorder such as vacancies on the other hand have zero-energy states themselves following a mechanism very similar to edges. Therefore, as long as their density is not too high, the zero energy peak in the density of states remains. We also verified numerically using our tight-binding model that the resulting energy shift of the bound states is small as long as the density of edge vacancies not too large. Thus, the large zero energy peak is robust against edge impurities.

\textit{Conclusions.-}
We have presented an effective  low-energy theory  to analyze odd-valley superconductivity in 2D materials that have a Fermi surface split into two valleys. This type of unconventional pairing is allowed by symmetry and involves an odd-parity pair potential that has an opposite sign in different valleys but remains isotropic within each valley. Our model predicts the existence of an almost flat band of edge states if the normal to the edge is such that the line of integration in k-space passes through both valleys. To confirm this prediction, we have also studied a tight-binding Hamiltonian for f-wave superconductivity in honeycomb lattices. Our findings showed massively degenerate edge states that appear as a pronounced zero energy peak in the density of states accessible through local spectroscopic techniques.

Our study has broad applicability to materials with hexagonal or triangular lattices, including graphene, NbSe, $\text{MoS}_{2}$ \cite{lu2015evidence}, nitrides \cite{yamanaka1996new,yamanaka1998superconductivity}, germanene \cite{xi2022superconductivity}, and silicene \cite{zhao2016rise}.
\section{Acknowledgements}
We would like to thank Stevan Nadj-Perge, F.~Guinea, M.A.~Cazalilla, A.A.~Golubov, A.~H.~MacDonald, A.~Vishwanath, S.~Suzuki for useful discussions.
T.K. and S.B.  acknowledge  financial support from Spanish MCIN/AEI/ 10.13039/501100011033 through project PID2020-114252GB-I00 (SPIRIT) and  TED2021-130292B-C42,
the Basque Government through grant IT-1591-22, and  European Union’s Horizon 2020 Research and Innovation Framework Programme under Grant No. 800923 (SUPERTED). I.V.T. acknowledges support by Grupos Consolidados UPV/EHU del Gobierno Vasco (Grant IT1453-22) and by the grant PID2020-112811GB-I00 funded by MCIN/AEI/10.13039/501100011033.
\nocite{*}
\bibliography{sources}




\title{Spectroscopic signature of spin triplet odd-valley  superconductivity in two-dimensional materials--Supplemental Material}
\author{T.H. Kokkeler\orcid{0000-0001-8681-3376}}
\email{tim.kokkeler@dipc.org}
\affiliation{Donostia International Physics Center (DIPC), 20018 Donostia--San Sebasti\'an, Spain}
\affiliation{University of Twente, 7522 NB Enschede, The Netherlands}

\author{Chunli Huang\orcid{0000-0002-0928-6266}}
\email{chunli.huang@uky.edu}
\affiliation{Department of Physics and Astronomy, University of Kentucky, Lexington, Kentucky 40506-0055, USA}

\author{F.S. Bergeret\orcid{0000-0001-6007-4878}}
\email{fs.bergeret@csic.es}
\affiliation{Centro de F\'isica de Materiales (CFM-MPC) Centro Mixto CSIC-UPV/EHU, E-20018 Donostia-San Sebasti\'an,  Spain}
\affiliation{Donostia International Physics Center (DIPC), 20018 Donostia--San Sebasti\'an, Spain}

\author{I. V. Tokatly\orcid{0000-0001-6288-0689}}
\email{ilya.tokatly@ehu.es}
\affiliation{IKERBASQUE, Basque Foundation for Science, 48009 Bilbao, Basque Country, Spain}
\affiliation{Donostia International Physics Center (DIPC), 20018 Donostia--San Sebasti\'an, Spain}
\affiliation{Nano-Bio Spectroscopy Group and European Theoretical Spectroscopy Facility (ETSF), Departamento de Polímeros 
y Materiales Avanzados: Física, Química y Tecnología, Universidad del País Vasco, 20018 Donostia-San Sebastián, 
Basque Country, Spain}

\maketitle
\appendix
\onecolumngrid\
\newpage
\section{Hamiltonian}\label{sec:Hamiltonian}
In the main text,  we use the low-energy Hamiltonian given by Eq. 1 of the main body. In this section, we elaborate on the structure of this Hamiltonian and show how this Hamiltonian can be obtained under general assumptions. We assume that the superconductivity is not nematic, that is, the crystal symmetry is respected and there are only two non-equivalent Dirac points. We consider zero-momentum Cooper pairs and triplet superconductivity. Lastly, we assume that the pair potential varies slowly with momentum on the scale of the Fermi momentum.

We introduce the following eight-component spinors, indexed by  $\vec{k}$ which satisfies $|\vec{k}|\ll|\vec{K}|$:
\begin{align}
\Psi = 
\begin{bmatrix}    
    \psi_{\alpha,\vec{k}\uparrow}\\\psi_{\alpha,\vec{k}\downarrow}\\\psi_{\alpha,-\vec{k}\downarrow}^{\dagger}\\-\psi_{\alpha,-\vec{k}\uparrow}^{\dagger}
\end{bmatrix}\otimes |u_{\alpha,\vec{K}+\delta\vec{k}}\rangle,
\end{align}
where $|u_{\vec{K}+\delta\vec{k}}\rangle$ is the Bloch function at momentum $\vec{k}$. 
The single particle energy is described by $\xi_{\vec{k}}$ for electrons, and thus $-\xi_{\vec{k}}$ for holes. We assume that the superconductivity is carried by zero-momentum Cooper pairs, so that $\langle\psi_{(\vec{K}+\delta\vec{k})\sigma}\psi_{(\vec{K}+\delta\vec{k})\sigma'}\rangle$ must vanish for all $\sigma,\sigma'$, but $\langle\psi_{(\vec{K}+\delta\vec{k})\sigma\psi_{-(\vec{K}+\delta\vec{k})\sigma'}}$ can be nonzero.
We consider general spin-triplet pairing $\Delta\vec{d}(\vec{k})\cdot\sigma$. This pairing must be odd-parity so that the Hamiltonian can be written as 
\begin{align}
    H(\vec{k}) &= \sum_{\alpha}        H_{\alpha}\otimes|u_{\alpha,\vec{k}}\rangle\langle u_{\alpha,\vec{k}}|,\\
    H_{\alpha}(\vec{k})&=\begin{bmatrix}
        \xi_{\alpha,\vec{k}}&\pm\Delta_{\alpha,\vec{k}}\\
        \pm\Delta_{\alpha,\vec{k}}(\vec{k})&-\xi_{\alpha,\vec{k}}
    \end{bmatrix}.
\end{align}
With this we arrive at the model in Eq. 1 of the main body.
\section{Tight-binding}
In this section, we analyze a tight-binding model with two different 
types of pairings: nearest-neighbor and next-nearest-neighbor pairings. 
 The low-energy model obtained   taking the limit $t\gg\mu\gg\Delta_{0}$, where $t$ is the hopping parameter, $\mu$ is the chemical potential and $\Delta_{0}$ is the energy of the superconducting pairing of two electrons on different sites. We show that nearest-neighbor pairing  produces  only interband superconductivity, but next-nearest-neighbour hopping produces the intraband superconductivity discussed in this work.
In this section the following definitions are used:
\begin{itemize}
    \item $\mu$, the chemical potential
    \item $t$, the hopping parameter
    \item $\Delta_{0}$ the pairing potential between two lattice sites
    \item $a_{i},b_{i}$ annihilation operators for the two different sublattices in real space.
    \item $a_{\vec{k}},b_{\vec{k}}$ annihilation operators for the two different sublattices in momentum space.
    \item $\alpha_{\vec{k}},\beta_{\vec{k}}$ annihilation operators in momentum space for conduction and valence band particles respectively.
    \item The distance between nearest neighbours $a$ is set to 1.
    \item $\Delta = \Delta_{0}s(\vec{K}) = \frac{3\sqrt{3}}{2}\Delta_{0}$.
    \item $k_{F} = \frac{2\mu}{3t}$ is the Fermi momentum.
    \item $V$ is strength of impurities, for edges we take $V\xrightarrow{}\infty$.
    \item Quantities denoted by $\psi$ are wavefunctions. They have subscripts merely to number them.
    \item $\Theta$ denotes the Heaviside function.
\end{itemize}

\subsection{Choice of pair potential}\label{sec:ChoicePP}
We consider two types of pair potentials, one with nearest neighbour hopping, and one with next nearest neighbour hopping. We assume that the pair potential has the same symmetry as the underlying lattice with only two inequivalent 
$\vec{K}$-points. This leaves only one option for odd-parity superconductivity in each case.

A tight-binding Hamiltonian for graphene with nearest neighbour pairing is 
\begin{align}\label{eq:NNSC}
    H_{1} &= -\mu \sum_{i}(a_{i}^{\dagger}a_{i}+b_{i}^{\dagger}b_{i})+t\sum_{i}\sum_{j = 1,2,3}a_{i}^{\dagger}b_{i+\delta_{j}}\nonumber\\&+i\frac{\Delta_{0}}{2}\sum_{i}\sum_{j = 1,2,3}(a_{i}^{\dagger}b_{i+\delta_{j}}^{\dagger}-b_{i+\delta_{j}}^{\dagger}a_{i}^{\dagger})
   +\text{h.c.}, 
\end{align}
where $a,b$ denote annihilation operators on the two nonequivalent sites in the unit cell,  $\delta_{1} = (0,-1),\delta_{2} = \frac{1}{2}(\sqrt{3},1),\delta_{3} = \frac{1}{2}(-\sqrt{3},1)$ are vectors between nearest neighbours.
A 2D Fourier transform over real space results in
\begin{align}
    H_{1} = \sum_{\vec{k}}-\mu (a_{\vec{k}}^{\dagger}a_{\vec{k}}+b_{\vec{k}}^{\dagger}b_{\vec{k}}) +t f(\vec{k}) a_{\vec{k}}^{\dagger}b_{\vec{k}}+i\frac{\Delta_{0}}{2}\bigg(f(\vec{k})a_{\vec{k}}^{\dagger}b_{-k}^{\dagger}-f^{*}(k)b_{\vec{k}}^{\dagger}a_{-k}^{\dagger}\bigg)+\text{h.c.},
\end{align}
where  the momentum dependence of the gap is given by  form factor
\begin{align}
    f(\vec{k}) &= e^{ik_{y}}+2e^{-\frac{i}{2}k_{y}}\cos(\frac{\sqrt{3}}{2}k_{x}).
\end{align}

A tight-binding Hamiltonian for graphene with next-nearest neighbour pairing is 
\begin{align}\label{eq:NNNSC}
    H_{2} &= -\mu \sum_{i}(a_{i}^{\dagger}a_{i}+b_{i}^{\dagger}b_{i})+t\sum_{i}\sum_{j = 
    1,2,3}a_{i}^{\dagger}b_{i+\delta_{j}}\nonumber\\&+i\frac{\Delta_{0}}{2}\sum_{i}\sum_{j = 1,2,3}(a_{i+\chi_{j}}^{\dagger}a_{i}^{\dagger}-a_{i-\chi_{j}}^{\dagger}a_{i}^{\dagger})\nonumber\\&+i\frac{\Delta_{0}}{2}\sum_{i}\sum_{j = 1,2,3}(b_{i+\chi_{j}}^{\dagger}b_{i}^{\dagger}-b_{i-\chi_{j}}^{\dagger}b_{i}^{\dagger})
   +\text{h.c.}, 
\end{align}
where $\chi_{1} = \delta_{3}-\delta_{2},\chi_{2} = \delta_{1}-\delta_{3},\chi_{3} = \delta_{2}-\delta_{1}$ are vectors between next-nearest neighbours. This pairing is the f-wave pairing introduced in \cite{pangburn2022superconductivityI}.
A Fourier transform over real space results in
\begin{align}
    H_{2} = \sum_{\vec{k}}-\mu (a_{\vec{k}}^{\dagger}a_{\vec{k}}+b_{\vec{k}}^{\dagger}b_{\vec{k}}) +t f(\vec{k}) a_{\vec{k}}^{\dagger}b_{\vec{k}}+\Delta_{0} s(\vec{k})(a_{\vec{k}}^{\dagger}a_{-k}^{\dagger}+b_{\vec{k}}^{\dagger}b_{-k}^{\dagger})+\text{h.c.},
\end{align}
where the momentum dependence of the gap is given by  form factor 
\begin{align}
    s(\vec{k}) &= \sin(\frac{\sqrt{3}}{2}k_{x}-\frac{3}{2}k_{y}
    )+\sin(\frac{\sqrt{3}}{2}k_{x}-\frac{3}{2}k_{y}
    )-\sin(\sqrt{3}k_{x}).
\end{align}

First, we consider the hopping terms, which are the same for both types of pairing. We define $f(\vec{k}) = F(\vec{k})e^{i\theta(\vec{k})}$, where $F,\theta$ are real variables, $F$ being positive. Since by symmetry $f$ satisfies $f(\vec{k}) = f^{*}(-\vec{k})$ its magnitude and phase satisfy $F(\vec{k}) = F(-\vec{k})$ and $\theta(\vec{k}) = -\theta(-\vec{k})$. The hopping terms are diagonalized by \begin{align}
    \alpha_{\vec{k}} &= \frac{1}{\sqrt{2}}(e^{-i\frac{\theta}{2}}a_{\vec{k}}+e^{i\frac{\theta}{2}}b_{\vec{k}}),\\
    \beta_{\vec{k}} &= \frac{1}{\sqrt{2}}(e^{-i\frac{\theta}{2}}a_{\vec{k}}-e^{i\frac{\theta}{2}}b_{\vec{k}}).
\end{align}
In terms of $\alpha_{\vec{k}}$ and $\beta_{\vec{k}}$ the Hamiltonian for nearest neighbour pairing is
\begin{align}
    H_{1} = -(\mu-tF(\vec{k}))\alpha_{\vec{k}}^{\dagger}\alpha_{\vec{k}}-(tF(\vec{k})+\mu)\beta_{\vec{k}}^{\dagger}\beta_{\vec{k}}+\Delta_{0}F(\vec{k})(\alpha_{\vec{k}}^{\dagger}\beta_{-\vec{k}}^{\dagger}-\beta_{\vec{k}}^{\dagger}\alpha_{-\vec{k}}^{\dagger})+\text{h.c.},
\end{align}
and the Hamiltonian for next nearest neighbour pairing is
\begin{align}
    H_{2} = -(\mu-tF(\vec{k}))\alpha_{\vec{k}}^{\dagger}\alpha_{\vec{k}}-(tF(\vec{k})+\mu)\beta_{\vec{k}}^{\dagger}\beta_{\vec{k}}+\Delta_{0} s(\vec{k})(\alpha_{\vec{k}}^{\dagger}\alpha_{-\vec{k}}^{\dagger}+\beta_{\vec{k}}^{\dagger}\beta_{-\vec{k}}^{\dagger})+\text{h.c.}.
\end{align}
Thus, nearest neighbour pairing gives inter-band pairing, $\alpha_{\vec{k}}^{\dagger}\beta_{-\vec{k}}^{\dagger}$, whereas next nearest neighbour pairing gives intra-band pairing, $\alpha_{\vec{k}}^{\dagger}\alpha_{-\vec{k}}^{\dagger}$ and $\beta_{\vec{k}}^{\dagger}\beta_{-\vec{k}}^{\dagger}$. Since the Fermi level only crosses one of the two bands and $\Delta\ll\mu$, interband pairing is heavily suppressed and only next nearest neighbour pairing should be considered. Since $s(\vec{K}) = -s(-\vec{K})\neq 0$, we may conclude that our tight-binding model with next nearest neighbour pairing in the limit $t\gg\mu\gg\Delta_{0}$ is described by the effective low-energy model described in the main body by Eq. (1). In the next section, we discuss the dispersion of the bound states in this tight-binding model.
\subsection{Bound states in the tight-binding model}
The Hamiltonian with next nearest neighbour pairing can be written in matrix form:
\begin{align}\label{eq:IntraKspace}
    H = \begin{bmatrix}
        -\mu& tf(\vec{k})&\Delta_{0} s(\vec{k})&0\\tf^{*}(k)&-\mu&0&\Delta_{0} s(\vec{k})\\\Delta_{0}s(\vec{k})&0&\mu&-tf(\vec{k})\\0&\Delta_{0}s(\vec{k})&-tf^{*}(k)&\mu
    \end{bmatrix}.
\end{align}

The band touchings are at the $\pm \vec{K}$-points, that is, at $\pm(\frac{4\pi}{3\sqrt{3}},0)$.
It is instructive to write this in a form with projections on the two bands of the system.
Defining $f(\vec{k}) = F(\vec{k})e^{i\phi(\vec{k})}$, 
\begin{align}
    G(k_{x},k_{y}) &= \frac{1}{(tF(\vec{k})-\mu)^{2}+(\Delta_{0} s(\vec{k}))^{2}-\omega^{2}}\begin{bmatrix}
        \omega-tF(\vec{k})+\mu&\Delta_{0} s(\vec{k})\\\Delta_{0} s(\vec{k})&\omega-\mu+tF(\vec{k})
    \end{bmatrix}\otimes\frac{1}{2}(1+\cos\phi\rho_{x}-\sin\phi\rho_{y})\nonumber\\&+\frac{1}{(tF(\vec{k})+\mu)^{2}+(\Delta_{0} s(\vec{k}))^{2}-\omega^{2}}\begin{bmatrix}
        \omega+tF(\vec{k})+\mu&\Delta_{0} s(\vec{k})\\\Delta_{0} s(\vec{k})&\omega-\mu-tF(\vec{k})
    \end{bmatrix}\otimes\frac{1}{2}(1-\cos\phi\rho_{x}+\sin\phi\rho_{y}); , 
\end{align}
where $\otimes$ denotes the Kronecker product and $\rho_{x,y}$ are the first and second Pauli matrices in sublattice space. The first term corresponds to the conduction band, the second to the valence band.

We first consider $k_{y} = 0$. In that case $f(\vec{k})$ is real, i.e. $\phi\in\{0,\pi\}$, and therefore Eq. B11 can be written as
\begin{align}
    H(k_{y} = 0) &= \begin{bmatrix}
        tf(\vec{k})-\mu&\Delta_{0} s(\vec{k})\\\Delta_{0} s(\vec{k})&\mu-tf(\vec{k})
    \end{bmatrix}\otimes\frac{1}{2}(1+\rho_{x})\nonumber\\&+\begin{bmatrix}
        -tf(\vec{k})-\mu&\Delta_{0} s(\vec{k})\\\Delta_{0} s(\vec{k})&\mu+tf(\vec{k})
    \end{bmatrix}\otimes\frac{1}{2}(1-\rho_{x}).
\end{align}
Thus,
\begin{align}\label{eq:Gky0}
    G(k_{x},k_{y} = 0) &= \frac{-1}{(tf(\vec{k})-\mu)^{2}+(\Delta_{0} s(\vec{k}))^{2}-\omega^{2}}\begin{bmatrix}
        \omega-tf(\vec{k})+\mu&\Delta_{0} s(\vec{k})\\\Delta_{0} s(\vec{k})&\omega-\mu+tf(\vec{k})
    \end{bmatrix}\otimes\frac{1}{2}(1+\rho_{x})\nonumber\\&+\frac{-1}{(tf(\vec{k})+\mu)^{2}+(\Delta_{0} s(\vec{k}))^{2}-\omega^{2}}\begin{bmatrix}
        \omega+tf(\vec{k})+\mu&\Delta_{0} s(\vec{k})\\\Delta_{0} s(\vec{k})&\omega-\mu-tf(\vec{k})
    \end{bmatrix}\otimes\frac{1}{2}(1-\rho_{x}).
\end{align}
Although $f(\vec{k})$ is real, it may still have either positive or negative sign. If $f(\vec{k})>0$ then $\phi = 0$ and thus the first term in Eq. B14 corresponds to the conduction band and the second term to the valence band, and if $f(\vec{k})<0$, then $\phi = \pi$ and the first term corresponds to the valence band instead. Thus, both terms in Eq. B14 contribute to the final result and should be evaluated. The Fourier integral should be taken over the projected Brillouin zone.
We define
\begin{align}
    G(x,k_{y} = 0) = \int_{-\frac{2\pi}{\sqrt{3}}}^{\frac{2\pi}{\sqrt{3}}}G(k_{x},k_{y} = 0)e^{ik_{x}x} dk_{x}.
\end{align}
This integral can be calculated analytically. 
In the coming part we use, for clarity of notation $\Delta = \Delta_{0}s(\vec{k})$.
A contour is taken that runs over the real line on between $k_{x} = \pm \frac{2\pi}{\sqrt{3}}$ and lines parallel to the imaginary axis at $\text{Re}(k_{x}) = \pm \frac{2\pi}{\sqrt{3}}$, closing it between $k_{x} = i\alpha\pm \frac{2\pi}{\sqrt{3}}$, where $\alpha$ is taken very large, positive if $x>0$, and negative if $x<0$. 
Now, lattice points exist for $x\in\{n\frac{\sqrt{3}}{2}\}$. For such points for each $\kappa$ it holds that $e^{i(\pm \frac{2\pi}{\sqrt{3}}+i\kappa)x} = -e^{-\kappa x}$, regardless of the choice of sign. 
This implies that the contributions of the lines of the lines parallel to the imaginary axis cancel out. Namely, the orientation of the lines is opposite while $G(k_{x},k_{y})$ is the same on both lines. The contribution of the top line is of order $\frac{e^{|\alpha|}}{\cosh^{2}\alpha}\approx e^{-|\alpha|}$ and thus vanishes as $|\alpha|\xrightarrow{}\infty$. Thus, the integral can be evaluated by a summation over the residues at the poles. For $t\gg\mu\gg\Delta_{0}$ these poles are close to the $\pm \vec{K}$-points.
Close to the $\pm \vec{K}$ points the functions $f$ and $s$ satisfy
\begin{align}
    f(\pm\vec{K}+\vec{p})\approx\frac{3}{2}(\pm p_{x}+ip_{y}),\label{eq:FclosetoK}\\
    s(\vec{k})\approx s(\pm\vec{K}) = \pm \frac{3\sqrt{3}}{2}.\label{eq:SclosetoK}
\end{align}
where $p_{x,y} = tk_{x,y}$. 
The first term in Eq. (B14) has poles at $k_{x} = K_{x}+\frac{2\mu}{3t}\pm 
i\frac{1}{t}\sqrt{\Delta^{2}-\omega^{2}}$ and $k_{x} = -K_{x}-\frac{2\mu}{3t}\pm i\frac{1}{t}\sqrt{\Delta^{2}-\omega^{2}}$. Similarly, for the second term in Eq. (B14) the poles are at 
$k_{x} = K_{x}-\frac{2\mu}{3t}\pm i\frac{1}{t}\sqrt{\Delta^{2}-\omega^{2}}$ and $k_{x} = -K_{x}+\frac{2\mu}{3t}\pm i\frac{1}{t}\sqrt{\Delta^{2}-\omega^{2}}$. Evaluating the residues at the poles with $\text{Im}(k_{x})>0$ if $x>0$ and the  residues with $\text{Im}(k_{x})<0$ for $x<0$ it follows that 
the Green's function for $k_{y} = 0$ is given by
\begin{align}
    \frac{4}{3t\pi i}G(x,k_{y} = 0)\approx
    \Bigg(\frac{1}{i\sqrt{\Delta^{2}-\omega^{2}}}\begin{bmatrix}
        \omega\cos{\Big((K_{x}+k_{F})x\Big)}&i\Delta\sin{\Big((K_{x}+k_{F})x\Big)}\\i\Delta\sin{\Big((K_{x}+k_{F})x\Big)}&\omega\cos{\Big((K_{x}+k_{F})x\Big)}
    \end{bmatrix}-i\tau_{3}\sin{\Big((K_{x}+k_{F})x\Big)}\text{sign}(x)\Bigg)e^{-\kappa|x|}\otimes\frac{1}{2}(1+\rho_{x})\nonumber\\
    +\Bigg(\frac{1}{i\sqrt{\Delta^{2}-\omega^{2}}}\begin{bmatrix}
        \omega\cos{\Big((K_{x}-k_{F})x\Big)}&i\Delta\sin{\Big((K_{x}-k_{F})x\Big)}\\i\Delta\sin{\Big((K_{x}-k_{F})x\Big)}&\omega\cos{\Big((K_{x}-k_{F})x\Big)}
    \end{bmatrix}+i\tau_{3}\sin\Big((K_{x}-k_{F})x\Big)\text{sign}(x)\Bigg)e^{-\kappa|x|}\otimes\frac{1}{2}(1-\rho_{x}),
\end{align}
where $k_{F} = \frac{2\mu}{3t}$. The bound states are at $\omega = \frac{\Delta}{\sqrt{1+V^{2}}}$, and for finite $V$ the spinors are given by
\begin{align}
    \psi_{1} = \Bigg(
        \frac{1}{iV}\cos{\Big((K_{x}+k_{F})x\Big)}\begin{bmatrix}
           1\\1\\0\\0 
        \end{bmatrix}+\sin{\Big((K_{x}+k_{F})x\Big)}\begin{bmatrix}
           -\text{sign(x)}\\-\text{sign}(x)\\\frac{\sqrt{1+V^{2}}}{iV}\\\frac{\sqrt{1+V^{2}}}{iV} 
        \end{bmatrix}
    \Bigg)e^{-\kappa|x|},\\
    \psi_{2} = \Bigg(
        \frac{1}{iV}\cos{\Big((K_{x}+k_{F})x\Big)}\begin{bmatrix}
           0\\0\\1\\1 
        \end{bmatrix}+\sin{\Big((K_{x}+k_{F})x\Big)}\begin{bmatrix}
           \frac{\sqrt{1+V^{2}}}{iV}\\\frac{\sqrt{1+V^{2}}}{iV} \\\text{sign(x)}\\\text{sign}(x)
        \end{bmatrix}
    \Bigg)e^{-\kappa|x|},\\
    \psi_{3} = \Bigg(
        \frac{1}{iV}\cos{\Big((K_{x}-k_{F})x\Big)}\begin{bmatrix}
           1\\-1\\0\\0 
        \end{bmatrix}+\sin{\Big((K_{x}-k_{F})x\Big)}\begin{bmatrix}
           \text{sign(x)}\\-\text{sign}(x)\\\frac{\sqrt{1+V^{2}}}{iV}\\-\frac{\sqrt{1+V^{2}}}{iV} 
        \end{bmatrix}
    \Bigg)e^{-\kappa|x|},\\
    \psi_{4} = \Bigg(
        \frac{1}{iV}\cos{\Big((K_{x}-k_{F})x\Big)}\begin{bmatrix}
           0\\0\\1\\-1 
        \end{bmatrix}+\sin{\Big((K_{x}-k_{F})x\Big)}\begin{bmatrix}
           \frac{\sqrt{1+V^{2}}}{iV}\\-\frac{\sqrt{1+V^{2}}}{iV} \\-\text{sign(x)}\\\text{sign}(x)
        \end{bmatrix}
    \Bigg)e^{-\kappa|x|}.
\end{align}
As $V\xrightarrow{}\infty$ it is convenient to consider $\psi_{A\pm} = \psi_{1}\pm i\psi_{2}$ and $\psi_{B\pm} = \psi_{3}\pm i\psi_{4}$. For $V\xrightarrow{}\infty$ these become
\begin{align}
    \psi_{A+} &= \sin{\Big((K_{x}+k_{F})x\Big)}\begin{bmatrix}
     1\\1\\-i\\-i   
    \end{bmatrix}e^{\kappa x}\Theta(-x),\\
    \psi_{A-} &= -\sin{\Big((K_{x}+k_{F})x\Big)}\begin{bmatrix}
     1\\1\\i\\i   
    \end{bmatrix}e^{-\kappa x}\Theta(x),\\
    \psi_{B+} &= -\sin{\Big((K_{x}-k_{F})x\Big)}\begin{bmatrix}
     1\\-1\\-i\\i   
    \end{bmatrix}e^{-\kappa x}\Theta(x),\\
    \psi_{B-} &= \sin{\Big((K_{x}-k_{F})x\Big)}\begin{bmatrix}
     1\\-1\\i\\-i   
    \end{bmatrix}e^{\kappa x}\Theta(-x).
\end{align}
Thus, there are four bound states, two on each side of the potential barrier. These states differ  in their sublattice and  Nambu space structure. In the main text this situation corresponds to the Nambu spinors at the two poles being orthogonal, $\langle u_{\vec{k}_{1}}|u_{\vec{k_{2}}}\rangle = 0$.
\subsubsection{Nonzero $k_{y}$}
We now consider the general case  $k_{y}\neq 0$. Define $\theta$ such that $k_{y} = k_{F}\sin\theta$. 
From Eq. B16 we find that close to the $K$-point $F\approx\frac{3}{2}\sqrt{p_{x}^{2}+p_{y}^{2}}$, $e^{i\phi} = \frac{p_{x}+ip_{y}}{\sqrt{p_{x}^{2}+p_{y}^{2}}}$ and $s(\vec{k})\approx\frac{3\sqrt{3}}{2}$, and close to the $-K$ point $F\approx\frac{3}{2}\sqrt{p_{x}^{2}+p_{y}^{2}}$, $e^{i\phi} = \frac{-p_{x}+ip_{y}}{\sqrt{p_{x}^{2}+p_{y}^{2}}}$ and $s(\vec{k})\approx-\frac{3\sqrt{3}}{2}$. Thus,
the poles are located at $\sqrt{p_{x}^{2}+p_{y}^{2}}\approx p_{F}$, i.e. $p_{x}\approx \pm p_{F}\cos\theta$. Expanding $p_{x} = p_{F}\cos\theta+q_{x}$, it follows that $q_{x} \approx \pm i\frac{1}{t\cos\theta}\sqrt{\Delta^{2}-\omega^{2}}$. Selecting the contribution from the poles similar to the case $k_{y} = 0$,
\begin{align}
    \frac{4\cos\theta}{3t\pi i}G(x,k_{y} = 0)&\approx
    \Bigg(\frac{1}{i\sqrt{\Delta^{2}-\omega^{2}}}\begin{bmatrix}
        \omega\cos{(K_{x}+k_{F}\cos\theta)x}&i\Delta\sin{(K_{x}+k_{F}\cos\theta)x}\\i\Delta\sin{(K_{x}+k_{F}\cos\theta)x}&\omega\cos{(K_{x}+k_{F}\cos\theta)x}
    \end{bmatrix}-i\tau_{3}\sin(K_{x}+k_{F}\cos\theta)x\text{sign}(x)\Bigg)e^{-\kappa|x|}\nonumber\\&\otimes\frac{1}{2}(1+\cos\theta\rho_{x}-\sin\theta\rho_{y})\nonumber\\
    &+\Bigg(\frac{1}{i\sqrt{\Delta^{2}-\omega^{2}}}\begin{bmatrix}
        \omega\cos{(K_{x}-k_{F}\cos\theta)x}&i\Delta\sin{(K_{x}-k_{F}\cos\theta)x}\\i\Delta\sin{(K_{x}-k_{F}\cos\theta)x}&\omega\cos{(K_{x}-k_{F}\cos\theta)x}
    \end{bmatrix}+\tau_{3}\sin(K_{x}-k_{F}\cos\theta)x\text{sign}(x)\Bigg)e^{-\kappa|x|}\nonumber\\&\otimes\frac{1}{2}(1-\cos\theta\rho_{x}-\sin\theta\rho_{y}).
\end{align}
The frequency of the bound states are unaltered compared to the previous case. An important difference is that the eigenfunctions of the projectors of first and second contribution are not orthogonal in this case. In the main body this situation corresponds to $\langle u_{\vec{k}_{1}}|u_{\vec{k_{2}}}\rangle\neq 0$. The bound states are

\begin{align}
    \psi_{1} &= \Bigg(
        \frac{1}{iV}\cos(K_{x}+k_{F}\cos\theta)x\begin{bmatrix}
           e^{i\frac{\theta}{2}}\\e^{-i\frac{\theta}{2}}\\0\\0 
        \end{bmatrix}+\sin(K_{x}+k_{F}\cos\theta)x\begin{bmatrix}
           -\text{sign(x)}(e^{i\frac{\theta}{2}})\\-\text{sign}(x)(e^{-i\frac{\theta}{2}})\\-\frac{\sqrt{1+V^{2}}}{iV}(e^{i\frac{\theta}{2}})\\-\frac{\sqrt{1+V^{2}}}{iV} (e^{-i\frac{\theta}{2}})
        \end{bmatrix}
    \nonumber\\
    &+\frac{i\sin\theta}{iV}\cos{(K_{x}-k_{F}\cos\theta)x}\begin{bmatrix}
        e^{-i\frac{\theta}{2}}\\-e^{i\frac{\theta}{2}}\\0\\0
    \end{bmatrix}+i\sin\theta\sin{(K_{x}-k_{F}\cos\theta)x}\begin{bmatrix}
        \text{sign}(x)e^{-i\frac{\theta}{2}}\\-\text{sign}(x)e^{i\frac{\theta}{2}}\\-\frac{\sqrt{1+V^{2}}}{iV}e^{-i\frac{\theta}{2}}\\\frac{\sqrt{1+V^{2}}}{iV}e^{i\frac{\theta}{2}}
    \end{bmatrix}\Bigg)e^{-\kappa|x|},
    \\
    \psi_{2} &= \Bigg(
        \frac{1}{iV}\cos(K_{x}+k_{F}\cos\theta)x\begin{bmatrix}
           0\\0\\e^{i\frac{\theta}{2}}\\e^{-i\frac{\theta}{2}} 
        \end{bmatrix}+\sin(K_{x}+k_{F}\cos\theta)x\begin{bmatrix}
           \frac{\sqrt{1+V^{2}}}{iV}e^{i\frac{\theta}{2}}\\\frac{\sqrt{1+V^{2}}}{iV} e^{-i\frac{\theta}{2}}\\-\text{sign(x)}e^{i\frac{\theta}{2}}\\-\text{sign}(x)e^{-i\frac{\theta}{2}}
        \end{bmatrix}\nonumber\\&+\frac{i\sin\theta}{iV}\cos(K_{x}-k_{F}\cos\theta)x\begin{bmatrix}
           0\\0\\e^{-i\frac{\theta}{2}}\\-e^{i\frac{\theta}{2}} 
        \end{bmatrix}+i\sin\theta\sin(K_{x}-k_{F}\cos\theta)x\begin{bmatrix}
           \frac{\sqrt{1+V^{2}}}{iV}e^{-i\frac{\theta}{2}}\\-\frac{\sqrt{1+V^{2}}}{iV} e^{i\frac{\theta}{2}}\\\text{sign(x)}e^{-i\frac{\theta}{2}}\\-\text{sign}(x)e^{i\frac{\theta}{2}}
        \end{bmatrix}
    \Bigg)e^{-\kappa|x|},\\
    \psi_{3} &= \Bigg(
        \frac{1}{iV}\cos(K_{x}-k_{F}\cos\theta)x\begin{bmatrix}
           (e^{-i\frac{\theta}{2}})\\-e^{i\frac{\theta}{2}}\\0\\0 
        \end{bmatrix}+\sin(K_{x}-k_{F}\cos\theta)x\begin{bmatrix}
           \text{sign}(x)e^{-i\frac{\theta}{2}}\\-\text{sign}(x)e^{i\frac{\theta}{2}}\\-\frac{\sqrt{1+V^{2}}}{iV}e^{-i\frac{\theta}{2}}\\\frac{\sqrt{1+V^{2}}}{iV} e^{i\frac{\theta}{2}}
        \end{bmatrix}\nonumber\\&-\frac{i\sin\theta}{iV}\cos(K_{x}+k_{F}\cos\theta)x\begin{bmatrix}
           e^{i\frac{\theta}{2}}\\e^{-i\frac{\theta}{2}}\\0\\0         \end{bmatrix}+i\sin\theta\sin(K_{x}+k_{F}\cos\theta)x\begin{bmatrix}
           \text{sign}(x)e^{i\frac{\theta}{2}}\\\text{sign}(x)e^{-i\frac{\theta}{2}}\\\frac{\sqrt{1+V^{2}}}{iV}e^{i\frac{\theta}{2}}\\\frac{\sqrt{1+V^{2}}}{iV} e^{-i\frac{\theta}{2}}
        \end{bmatrix}
    \Bigg)e^{-\kappa|x|},\\
    \psi_{4} &= \Bigg(
        \frac{1}{iV}\cos(K_{x}-k_{F}\cos\theta)x\begin{bmatrix}
           0\\0\\e^{-i\frac{\theta}{2}}\\-e^{i\frac{\theta}{2}} 
        \end{bmatrix}+\sin(K_{x}-k_{F}\cos\theta)x\begin{bmatrix}
           \frac{\sqrt{1+V^{2}}}{iV}e^{-i\frac{\theta}{2}}\\-\frac{\sqrt{1+V^{2}}}{iV}e^{i\frac{\theta}{2}} \\\text{sign(x)}e^{-i\frac{\theta}{2}}\\-\text{sign}(x)e^{\frac{\theta}{2}}
        \end{bmatrix}\nonumber\\&-\frac{i\sin\theta}{iV}\cos(K_{x}+k_{F}\cos\theta)x\begin{bmatrix}
           0\\0\\e^{i\frac{\theta}{2}}\\e^{-i\frac{\theta}{2}}
        \end{bmatrix}-i\sin\theta\sin(K_{x}+k_{F}\cos\theta)x\begin{bmatrix}
           \frac{\sqrt{1+V^{2}}}{iV}e^{i\frac{\theta}{2}}\\\frac{\sqrt{1+V^{2}}}{iV}e^{-i\frac{\theta}{2}} \\-\text{sign(x)}e^{i\frac{\theta}{2}}\\-\text{sign}(x)e^{-i\frac{\theta}{2}}
        \end{bmatrix}
    \Bigg)e^{-\kappa|x|}.
\end{align}
Again, we can make the combinations $\psi_{A\pm} = \psi_{1}\pm\psi_{2}$ and $\psi_{B\pm} = \psi_{3}\pm\psi_{4}$, and we obtain in the limit $V\xrightarrow{}\infty$
\begin{align}
    \psi_{A+} = \Bigg(\sin(K_{x}+k_{F}\cos\theta)x\begin{bmatrix}
     e^{i\frac{\theta}{2}}\\e^{-i\frac{\theta}{2}}\\ie^{i\frac{\theta}{2}}\\ie^{-i\frac{\theta}{2}}
    \end{bmatrix}+\sin(K_{x}-k_{F}\cos\theta)x\sin\theta\begin{bmatrix}
        -ie^{-i\frac{\theta}{2}}\\ie^{i\frac{\theta}{2}}\\-e^{-i\frac{\theta}{2}}\\e^{i\frac{\theta}{2}}
    \end{bmatrix}\Bigg)e^{-\kappa x}\Theta(x),\\
    \psi_{A-} = \Bigg(\sin(K_{x}+k_{F}\cos\theta)x\begin{bmatrix}
     e^{i\frac{\theta}{2}}\\e^{-i\frac{\theta}{2}}\\-ie^{i\frac{\theta}{2}}\\-ie^{-i\frac{\theta}{2}}
    \end{bmatrix}+\sin(K_{x}-k_{F}\cos\theta)x\sin\theta\begin{bmatrix}
        -ie^{-i\frac{\theta}{2}}\\ie^{i\frac{\theta}{2}}\\e^{-i\frac{\theta}{2}}\\-e^{i\frac{\theta}{2}}
    \end{bmatrix}\Bigg)e^{\kappa x}\Theta(-x),\\
    \psi_{B+} = \Bigg(\sin(K_{x}-k_{F}\cos\theta)x\begin{bmatrix}
        e^{-i\frac{\theta}{2}}\\-e^{i\frac{\theta}{2}}\\ie^{-i\frac{\theta}{2}}\\-ie^{i\frac{\theta}{2}}
\end{bmatrix}+\sin(K_{x}+k_{F}\cos\theta)x\sin\theta\begin{bmatrix}ie^{i\frac{\theta}{2}}\\ie^{-i\frac{\theta}{2}}\\e^{i\frac{\theta}{2}}\\e^{-i\frac{\theta}{2}}
\end{bmatrix}\Bigg)e^{-\kappa x}\Theta(x),\\
    \psi_{B-} = \Bigg(\sin(K_{x}-k_{F}\cos\theta)x\begin{bmatrix}
        e^{i\frac{\theta}{2}}\\-e^{-i\frac{\theta}{2}}\\ie^{i\frac{\theta}{2}}\\-ie^{-i\frac{\theta}{2}}
\end{bmatrix}+\sin(K_{x}+k_{F}\cos\theta)x\sin\theta\begin{bmatrix}ie^{i\frac{\theta}{2}}\\ie^{-i\frac{\theta}{2}}\\-e^{i\frac{\theta}{2}}\\-e^{-i\frac{\theta}{2}}
\end{bmatrix}\Bigg)e^{\kappa x}\Theta(-x).
\end{align}
Thus, also for $k_{y}\neq 0$ there exist two orthogonal states on each side of the line. In this case the states do not have a specific sublattice structure or a well defined single periodicity. 
\subsubsection{Higher order terms}\label{sec:HigherOrderTerms}
The analytical results presented in the main text and in the previous section,  are obtained in the limit $\mu\gg\Delta$ and$\mu\sin\theta\gg\Delta$. 
For $\theta\approx\frac{\pi}{2}$ the latter assumption is not valid, and the results of the tight-binding model do show a  dispersion for $k_{y}\approx k_{F}$. 
In this section we analytically derive that the dispersion for small $k_{y}$ is of order $\Delta^{2}/\mu$. Moreover, we show that for large $k_{y}$ the energy of the edge states approaches the conduction band.

The Green's function can be written as
\begin{align}
    G &= \frac{2}{X}(AG_{1}+BG_{2}),\nonumber\\
    X& = \left((tF(\vec{k}))^2+\mu^2+(\Delta_{0}s(\vec{k}))^{2}-\omega^{2}\right)^2-4\mu^{2}t^{2}F(\vec{k})^2,\nonumber\\
    A& = (tF(\vec{k}))^{2}+\mu^{2}+\Delta_{0}^{2}s(\vec{k})^{2}-\omega^{2},\nonumber\\
    B&= 2\mu, \nonumber\\
    G_{1}&=\begin{bmatrix}
        \omega+\mu&-tf&\Delta_{0}s(\vec{k})&0\\-t\Tilde{f}&\omega+\mu&0&\Delta_{0}s(\vec{k})\\
        \Delta_{0}s(\vec{k})&0&\omega-\mu&tf\\0&\Delta_{0}s(\vec{k})&t\Tilde{f}&\omega-\mu
    \end{bmatrix},\nonumber\\
    G_{2} & = \begin{bmatrix}
        -(tF)^{2}&(\omega+\mu)tf&0&\Delta_{0}s(\vec{k})tf\\
        (\omega+\mu)t\Tilde{f}&-(tF)^{2}&\Delta_{0}s(\vec{k})t\Tilde{f}&0\\
        0&\Delta s(\vec{k}) tf&(tF)^{2}&(\omega-\mu)tf\\\Delta_{0}s(\vec{k})t\Tilde{f}&0&(\omega-\mu)t\Tilde{f}&(tF)^{2}
    \end{bmatrix},\label{eq:Gdef}
\end{align}
where $f(\vec{k}) = e^{-ik_{y}}+2e^{i\frac{k_{y}}{2}}\cos\frac{\sqrt{3}}{2}k_{y}$ and $\Tilde{f}(k_{x},k_{y}) =f(k_{x},-k_{y})$. Because we assume $\frac{\mu}{t}\ll 1$ we may linearize around each $K$-point. Near the $K$-point, $f(\vec{k}-\vec{K})\approx \Tilde{t}(k_{x}+ik_{y})$, where $\Tilde{t} = \frac{3t}{2}$, and $\Delta \approx \Delta_{0}3\sqrt{3}$.
For simplicity of notation we denote the product of $\Tilde{t}$ and $k_{x,y}$ by $p_{x,y}$. 

In the first part we do assume that $\frac{\mu}{t}$ is small so that terms of higher order in $t/\mu$ can be ignored, the influence of those terms is discussed in Sec. B4. 
Under the assumption that $t\gg\mu$ the denominator is a fourth order polynomial, hence we may rewrite the Green's function $G$ near a $\vec{K}$-point as 
\begin{align}
    G = (AG_{1}+BG_{2})\prod_{i = 1}^{4}\frac{1}{p_{x}-p_{xi}},
    p_{x1} = \alpha+i\beta,p_{x2} = \alpha -i\beta, p_{x3} = -\alpha+i\beta, p_{x4} = -\alpha-i\beta,
\end{align}
where $\alpha$ and $\beta$ satisfy
\begin{align}
    \alpha^{2}-\beta^{2} &= \mu^{2}-p_{y}^{2}+\omega^{2}-\Delta^{2},\nonumber\\
    \alpha\beta &= \mu\sqrt{\Delta^{2}-\omega^{2}},\nonumber\\
    F(p_{xi}) &= \mu\pm i\sqrt{\Delta^{2}-\omega^{2}},
\end{align}
where the $+$ sign is to be used for poles $1$ and $4$, and the minus sign for poles 2 and 3.
Explicit expressions for $\alpha^{2}$ and $\beta^{2}$ are
\begin{align}
    \alpha^{2} &= \mu^{2}-k_{y}^{2}+\omega^{2}-\Delta^{2}+\sqrt{(\mu^{2}-k_{y}^{2}+\omega^{2}-\Delta^{2})^{2}+4\mu^{2}(\Delta^{2}-\omega^{2})},\\
    \beta^{2} &= k_{y}^{2}-\mu^{2}+\Delta^{2}-\omega^{2}+\sqrt{(\mu^{2}-k_{y}^{2}+\omega^{2}-\Delta^{2})^{2}+4\mu^{2}(\Delta^{2}-\omega^{2})}.
\end{align}
For the evaluation of the residues in the following we  need the quantities
\begin{align}
    \lim_{p_{x}\xrightarrow{}p_{xj}}(p_{x}-p_{xj})\prod_{i = 1}^{4}\frac{1}{p_{xj}-p_{xi}} = \prod_{i\neq j}\frac{1}{p_{xj}-p_{xi}}.
\end{align}
For $j = 1,4$ this quantity equals $\frac{1}{\alpha\beta(\alpha+i\beta)}$, for $j = 2,3$  it equals $\frac{1}{\alpha\beta(\alpha-i\beta)}$
For $x>0$ we have to evaluate poles 1 and 3. The diagonal elements in sublattice space of the blocks proportional diagonal in Nambu space, to be called the $(1,1)$-block and $(2,2)$-block, read
\begin{align}
    &\frac{2\mu}{\alpha\beta(\alpha+i\beta)}\left((\mu+i\sqrt{\Delta^{2}-\omega^{2}})(\omega+\mu)-(\mu+i\sqrt{\Delta^{2}-\omega^{2}})^{2}\right)e^{ip_{1}x}\nonumber\\
    &+\frac{2\mu}{\alpha\beta(\alpha-i\beta)}\left((\mu-i\sqrt{\Delta^{2}-\omega^{2}})(\omega+\mu)-(\mu-i\sqrt{\Delta^{2}-\omega^{2}})^{2}\right)e^{ip_{3}x}\nonumber\\
    &=\frac{2\mu}{\alpha\beta(\alpha^{2}+\beta^{2})}\Bigg((\alpha-i\beta)\left((\mu+i\sqrt{\Delta^{2}-\omega^{2}})(\omega-i\sqrt{\Delta^{2}-\omega^{2}})\right)e^{ip_{1}x}
    +(\alpha+i\beta)\left((\mu-i\sqrt{\Delta^{2}-\omega^{2}})(\omega+i\sqrt{\Delta^{2}-\omega^{2}})\right)e^{ip_{3}x}\Bigg).
\end{align}
In the limit $x\xrightarrow{}0$ this becomes
\begin{align}
    \frac{4\mu}{\alpha\beta(\alpha^{2}+\beta^{2})}\left(\alpha(\mu\omega+\Delta^{2}-\omega^{2})+\beta\sqrt{\Delta^{2}-\omega^{2}}(\omega-\mu)\right).
\end{align}
Similarly, the off-diagonal elements of this block at $x = 0$ are given by
\begin{align}
    &\frac{2\mu}{\alpha\beta(\alpha^{2}+\beta^{2})}\Bigg((\alpha-i\beta)(\omega+\mu-\mu-i\sqrt{\Delta^{2}-\omega^{2}})(\alpha+i\beta\pm ik_{y})+(\alpha+i\beta)(\omega+\mu-\mu+i\sqrt{\Delta^{2}-\omega^{2}})(-\alpha+i\beta \pm ik_{y})\Bigg)\nonumber\\
    &=\frac{4\mu}{\alpha\beta(\alpha^{2}+\beta^{2})}\Bigg(-(\alpha^{2}+\beta^{2})i\sqrt{\Delta^{2}-\omega^{2}}\pm ik_{y}(\alpha\omega-\beta\sqrt{\Delta^{2}-\omega^{2}})\Bigg).
\end{align}

Now consider the Green's function in the opposite valley. The Green's function is  similar in the two valleys. The only differences are that (i) $tf = -p_{x}+ip_{y}$ in the $-K$-valley whereas $tf = p_{x}+ip_{y}$ in the $K$-valley and (ii) $\Delta$ has opposite sign in the opposite valleys. This means that the diagonal elements of the Green's function are the same in both valleys, whereas in the off-diagonal elements of the $(1,1)$-block only those terms proportional to $p_{y}$ survive. The $(1,1)$-block in Nambu space thus reads
\begin{align}
    \frac{8\mu}{\alpha\beta(\alpha^{2}+\beta^{2})}\left(\alpha(\mu\omega+\Delta^{2}-\omega^{2})+\beta\sqrt{\Delta^{2}-\omega^{2}}(\omega-\mu)\mathbf{1}_{\rho}+k_{y}(\alpha\omega-\beta\sqrt{\Delta^{2}-\omega^{2}})\rho_{y}\right),
\end{align}
where $\mathbf{1}_{\rho}$ is the identity matrix in sublattice space. In an entirely similar way, the $(2,2)$-block in Nambu space is, multiplying the expression by -1 and then reversing the sign of $\omega$
\begin{align}
    \frac{8\mu}{\alpha\beta(\alpha^{2}+\beta^{2})}\left(\alpha(\mu\omega-\Delta^{2}+\omega^{2})+\beta\sqrt{\Delta^{2}-\omega^{2}}(\omega+\mu)\right)\mathbf{1}_{\rho}+k_{y}(\alpha\omega+\beta\sqrt{\Delta^{2}-\omega^{2}})\rho_{y}.
\end{align}
This closes the discussion of the blocks diagonal in Nambu space, to be called the $(1,2)$-block and $(2,1)$-block.
Next consider the terms that are off-diagonal in Nambu space. The procedure is similar. However, since the sum of the pair potentials in both valleys is zero, only the terms proportional to $p_{x}\Delta$ in the terms off-diagonal in sublattice space add up constructively for the two valleys, and the $\Delta$ and $p_{y}\Delta$ terms cancel out because they have opposite sign in the valleys. The surviving term is given by
\begin{align}
    \frac{2\mu}{\alpha\beta(\alpha^{2}+\beta^{2})}\Delta\left((\alpha-i\beta)(\alpha+i\beta)+(\alpha+i\beta)(-\alpha+i\beta)\right)) = 0,
\end{align}
that is, the terms off-diagonal in Nambu space vanish exactly.

Therefore, bound states occur if the determinant of one of either the $(1,1)$ or $(2,2)$-block vanishes to all orders in $\Delta/\mu$. Since the determinant of hole block can be obtained from the electron block by negation of $\omega$, it is sufficient to consider only the electron block and take into account that bound states always come in pairs at $\pm\omega$.
The determinant, ignoring the prefactor evaluates to
\begin{align}
    (\mu^{2}-k_{y}^{2})(\alpha\omega+\beta\sqrt{\Delta^{2}-\omega^{2}})^2+(\Delta^{2}-\omega^{2})(\alpha\sqrt{\Delta^{2}-\omega^{2}}+\beta\omega)^{2}+2\mu\sqrt{\Delta^{2}-\omega^{2}}((\alpha^{2}-\beta^{2})\omega\sqrt{\Delta^{2}-\omega^{2}}+\alpha\beta(2\omega^{2}-\Delta^{2})).
\end{align}
Evaluating this term and using that $\alpha^{2}-\beta^{2} = \mu^{2}+\omega^{2}-\Delta^{2}-p_{y}^{2}$ and $\alpha\beta = \mu\sqrt{\Delta^{2}-\omega^{2}}$, the expression can be written as
\begin{align}
    &(\mu^{2}+\omega^{2}-\Delta^{2}-k_{y}^{2})^{2}\omega^{2}+((\mu^{2}-k_{y}^{2})\beta^{2}+(\Delta^{2}-\omega^{2})\alpha^{2})\Delta^{2}+2\mu^{2}(2\omega^{2}-\Delta^{2})(\Delta^{2}-\omega^{2})\nonumber\\
    &= \left((\mu^{2}+\omega^{2}-\Delta^{2}-k_{y}^{2})^{2}+4\mu^{2}(\Delta^{2}-\omega^{2})\right)\omega^{2}+\left((\mu^{2}-k_{y}^{2})\beta^{2}+(\Delta^{2}-\omega^{2})\alpha^{2}-2\mu^{2}(\Delta^{2}-\omega^{2})\right)\Delta^{2}\nonumber\\
    &=\left((\alpha^{2}+\beta^{2})^{2}-(2\mu^{2}-\alpha^{2})\Delta^{2}\right)\omega^{2}+\left((\mu^{2}-k_{y}^{2})\beta^{2}+(\Delta^{2}-\omega^{2})\alpha^{2}-2\mu^{2}(\Delta^{2}-\omega^{2})\right)\Delta^{2}.
\end{align}
With this, we have an implicit expression for the bound state energies:
\begin{align}\label{eq:FirstExpBS}
    \omega^{2} = \frac{(2\mu^{2}-\alpha^{2})\Delta^{2}-(\mu^{2}-p_{y}^{2})\beta^{2}}{(\alpha^{2}+\beta^{2})^{2}+(2\mu^{2}-\alpha^{2})\Delta^{2}}\Delta^{2}.
\end{align}
However, note that $\alpha$ and $\beta$ do depend on $\omega$, so that this is an indirect expression.

From Eq. B50 the bound state energy can be found as a function of $k_{y}$. The results for $\Delta/\mu = 0.1$ and $\Delta/\mu = 0.05$ are shown in Supplemental Fig. 1. The results are in agreement with the results of the tight-binding model that for small $k_{y}$ the dispersion becomes smaller as the ratio $\Delta/\mu$ is decreased. Close to the $k_{y} = k_{F}$ the energy of the bound states increases, and for  $k_{y}>k_{F}$ the edge states approach the conduction band. 
\begin{figure*}
    \centering
    {\hspace*{-1.5em}\includegraphics[width = 8.6cm]{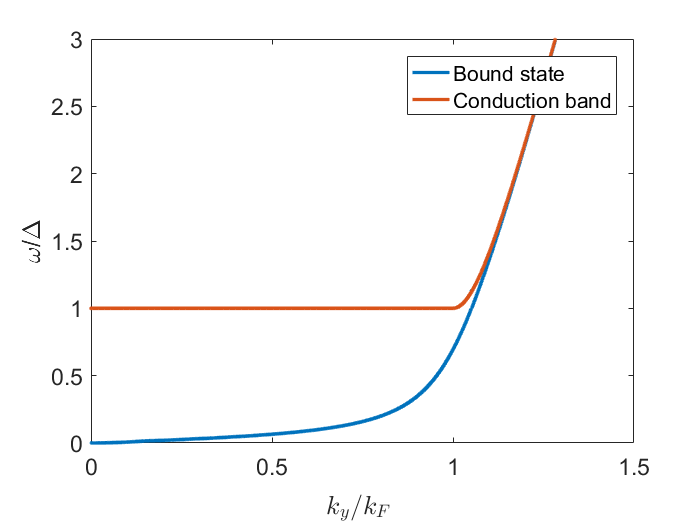}}
    \hfill
        {\hspace*{-2em}\includegraphics[width = 8.6cm]{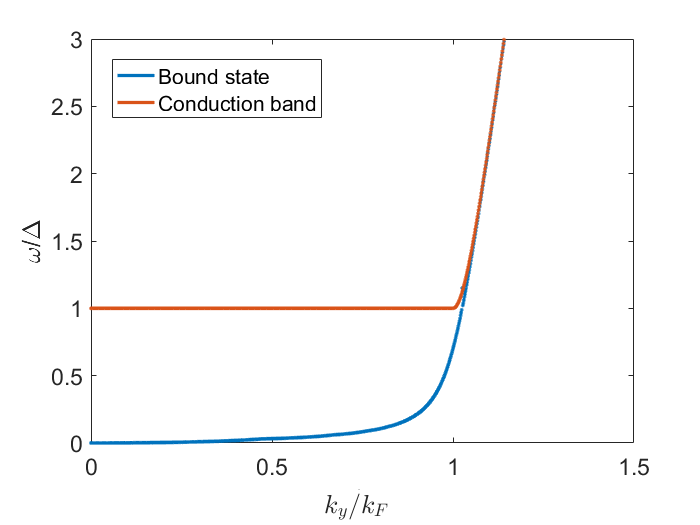}}
        \hfill
        {\hspace*{-2em}\includegraphics[width = 8.6cm]{figures/A.png}}\hfill
        {\hspace*{-2em}\includegraphics[width = 8.6cm]{figures/B.png}}
    \caption{The bound state energies as a function of $k_{y}$ for $\Delta/\mu = 0.1$ (a) and $\Delta/\mu = 0.05$ (b). For smaller $\Delta/\mu$ the dispersion around $k_{y} = 0$ becomes smaller. In both cases the conduction band is rapidly approached for $k_{y}>k_{F}$.}
    \label{fig:BoundStates}
\end{figure*}

In certain limits, direct analytical expressions can be obtained. First consider the case $p_{y}\ll\mu$, that is, $k_{y}\ll k_{F}$. Then $$\alpha^{2} = \mu^{2}-p_{y}^{2}-\frac{(\omega^{2}-\Delta^{2})p_{y}^{2}}{\mu^{2}}+O(p_{y}/\mu)^{4}$$ and $$\beta^{2}=\Delta^{2}-\omega^{2}-\frac{(\omega^{2}-\Delta^{2})p_{y}^{2}}{\mu^{2}}+O(p_{y}/\mu)^{4}.$$ Substituting this into Eq. B50, and ignoring terms of at least second order in $\Delta/\mu$
\begin{align}
    \omega^{2}&\approx\frac{(\mu^{2}+p_{y}^{2})\Delta^{2}-(\mu^{2}-p_{y}^{2})(\Delta^{2}-\omega^{2})+(\omega^{2}-\Delta^{2})p_{y}^{2}}{(\mu^{2}-p_{y}^{2}+\Delta^{2}-\omega^{2})^{2}+(\mu^{2}+p_{y}^{2})\Delta^{2}}\Delta^{2}\\
    &\approx\frac{\mu^{2}\omega^{2}+p_{y}^{2}\Delta^{2}}{(\mu^{2}-p_{y}^{2}+\Delta^{2}-\omega^{2})^{2}+(\mu^{2}+p_{y}^{2})\Delta^{2}}\Delta^{2}.
\end{align}
From here we see that to lowest order in $\frac{\Delta}{\mu}$

\begin{align}
    \omega = \frac{\Delta^{2}}{\mu^{2}}p_{y}.
\end{align}
In terms of $k_{y}$ this can be written as
\begin{align}
    \omega = \frac{\Delta^{2}}{\mu}\frac{k_{y}}{k_{F}}.
\end{align}
The above expressions explain the dispersion around $k_{y} = 0$ as observed in the tight-binding model and show that they are indeed higher order in $\Delta/\mu$.

The second case that we consider is the case in which $p_{y}$ is close to $\mu$, that is, $k_{y}$ is close to $k_{F}$. To be precise, we assume $\mu^{2}-p_{y}^{2} \ll\Delta^{2}-\omega^2$. Then,
\begin{align}
    \alpha^{2} &= \mu\sqrt{\Delta^{2}-\omega^{2}}+\frac{1}{2}(\mu^{2}-p_{y}^{2}+\omega^{2}-\Delta^{2})+O(\frac{\Delta^{3}}{\mu}),\\
    \beta^{2} &= \mu\sqrt{\Delta^{2}-\omega^{2}}-\frac{1}{2}(\mu^{2}-p_{y}^{2}+\omega^{2}-\Delta^{2})+O(\frac{\Delta}{\mu}).
\end{align}
It is enough to keep only the first order approximations,
\begin{align}\label{eq:ThirdCase}
    \omega^{2} &= \frac{2\mu^{2}\Delta^{2}-\mu\sqrt{\Delta^{2}-\omega^{2}}(\Delta^{2}+\mu^{2}-p_{y}^{2})}{\mu^{2}(6\Delta^{2}-4\omega^{2})-\mu\sqrt{\Delta^{2}-\omega^{2}}\Delta^{2}}\Delta^{2}\nonumber\\
    &=\frac{1-\frac{\sqrt{\Delta^{2}-\omega^{2}}}{2\mu}(1+\frac{\mu^{2}-p_{y}^{2}}{\Delta^{2}})}{3-2(\frac{\omega}{\Delta})^{2}-\frac{\sqrt{\Delta^{2}-\omega^{2}}}{2\mu}}\Delta^{2}.
\end{align}
To zeroth order approximation there are seemingly two solutions, $\omega = \Delta$ and $\omega = \Delta/\sqrt{2}$. The former of the two is in fact not really a solution, as $\alpha$ and $\beta$ vanish as well, so that $G(x = 0)$ has finite determinant.
Thus, the only solution that should be considered is $\omega = \Delta/\sqrt{2}+\Tilde{\omega}$ with $\Tilde{\omega}\ll\Delta$. Substituting this into Eq. B50 and ignoring any higher order term in $\Tilde{\omega}$, we find
\begin{align}
    \omega\approx\frac{\Delta}{\sqrt{2}}-\frac{\Delta^{2}}{8\mu}(1+2\frac{\mu^{2}-p_{y}^{2}}{\Delta^{2}}).
\end{align}
Note that since $p_{y}\approx\mu$ this implies that
\begin{align}
    \frac{d\omega}{dp_{y}}\approx \frac{1}{2}.
\end{align}
This implies that in a relatively small window, of order $\frac{\Delta}{\mu}k_{F}$ around $k_{F}$ the energy of the bound states changes by $\Delta$. Thus, for $\Delta/\mu\ll 1$ the bound state energy increases from almost zero to close to the conduction band in a narrow energy window.

A third interesting limit is the one in which $k_{y}\gg k_{F}$. 
Ignoring terms that are of order $\frac{1}{p_{y}^{2}}$, Eq. B50 becomes
\begin{align}
    \omega^{2} = \frac{2\mu^{2}\Delta^{2}+(p_{y}^{2}-\mu^{2})(p_{y}^{2}-\mu^{2}+\Delta^{2}-\omega^{2})}{2\mu^{2}\Delta^{2}+(p_{y}^{2}-\mu^{2}+\Delta^{2}-\omega^{2})(p_{y}^{2}-\mu^{2}+\Delta^{2}-\omega^{2})} \Delta^{2}.
\end{align}
One of the solutions is $\omega = \Delta$. However, for $\omega = \Delta$, it is found that the term $\alpha$ is actually vanishing as well, so that this is not a bound state. Thus, there are no bound states with $\omega\ll p_{y}$. Instead, Supplemental Fig. 1 suggests that the bound states are close to the conduction band, which is given by $E_{c} = (p_{y}-\mu)^{2}+\Delta^{2}$ for $p_{y}>\mu$. Approximating $\omega\approx p_{y}-\mu+\Tilde{\omega}$, where $|\Tilde{\omega}|\ll p_{y}-\mu$, we can  compute that
\begin{align}
    \alpha^{2} &\approx -(p_{y}-\mu)\mu-\frac{1}{2}\Delta^{2}+\frac{\Delta}{2}\sqrt{4\mu(p_{y}-\mu)+\Delta^{2}},\\
    \beta^{2} &\approx (p_{y}-\mu)\mu+\frac{1}{2}\Delta^{2}+\frac{\Delta}{2}\sqrt{4\mu(p_{y}-\mu)+\Delta^{2}}.
\end{align}
Note that $\alpha^{2}<0$. This is not a problem. It indicates that the two pairs of complex conjugate poles now both have vanishing real part and a different value for the imaginary part.
Substituting this into Eq. B50, and ignoring any terms that are not of lowest order in $\Delta^{2}$ or $\Tilde{\omega}$, the following equation is found:
\begin{align}
    \frac{\mu(p_{y}+\mu)(p_{y}-\mu)^{2}}{\mu k_{y}(-\Tilde{\omega}+2\Delta^{2})}\Delta^{2} = (p_{y}-\mu)^{2},
\end{align}
\begin{align}
    \Tilde{\omega} = 2\Delta^{2}-\Delta^{2}\frac{\mu(p_{y}+\mu)(p_{y}-\mu)^{2}}{\mu p_{y}(p_{y}-\mu)^{2}} = 2\Delta^{2}-\Delta^{2}(1+\frac{p_{y}}{\mu}) = \Delta^{2}(1-\frac{p_{y}}{\mu}).
\end{align}
Thus,
\begin{align}
    \omega^{2} &\approx (p_{y}-\mu)^{2} +\Delta^{2}(1-\frac{\mu}{p_{y}}) = E_{c}^{2}-\frac{\mu}{p_{y}}\Delta^{2},\nonumber\\
    \omega&\approx E_{c}-\frac{\mu}{p_{y}}\frac{\Delta^{2}}{E_{c}}.
\end{align}
This shows that the bound state approaches the conduction band quickly as $k_{y}\gg\mu$.
With this, we have analytically understood all main features of Eq. B50, explaining the influence of a finite $\Delta/\mu$.
\subsection{Localization length of edge states }\label{sec:Localization}
An important feature of edge states is their localization,
\begin{align}
    \langle x\rangle = \frac{1}{a}\int x|\psi|^{2}dx \left(\int |\psi|^{2}dx\right)^{-1},
\end{align}
where $a$ is the distance between two carbon atoms, here set to 1.
This quantity determines how far opposite edges must be apart to have small interaction between the two. For small $k_{y}$ it holds that $\alpha \approx\mu^{2}-p_{y}^{2}>0$ and $\beta^{2} = \Delta^{2}-\omega^{2}>0$, and therefore
\begin{align}
    \langle x\rangle\sim\frac{1}{\text{Im}(k_{x})} = \frac{3}{2}t(\Delta^{2}-\omega^{2})^{-\frac{1}{2}}\approx\frac{3}{2}t\Delta^{-1},
\end{align}
Thus, the edge states are localized on the order of a superconducting coherence length. For $k\approx k_{F}$, we have
$\alpha^{2}\approx\beta^{2} \approx(\frac{3}{2}t)^{-1}\mu\sqrt{\Delta^{2}-\omega^{2}}$, so
\begin{align}
    \langle x\rangle\sim\frac{1}{\text{Im}(k_{x})} = \frac{3}{2}t(\Delta^{2}-\omega^{2})^{-\frac{1}{4}}\mu^{-\frac{1}{2}}\approx\frac{3}{4}t(\Delta\mu)^{-\frac{1}{2}}.
\end{align}
Since $\mu\gg\Delta$, these states are more localized than the states near $k_{y} = 0$. If $k_{y}\gg k_{F}$ we have that
$\beta^{2} \approx -\alpha^{2} \approx (k_{y}-\mu)\mu$. Thus, $\alpha^{2}<0$, which means that the four solutions for $k_{x}$ are all purely imaginary, and for $\alpha\approx i\beta$,
\begin{align}
    \text{Im}(k_{x}) =  \pm 2\beta  = (\frac{3}{2}t)^{-1}\sqrt{(k_{y}-\mu)\mu},
\end{align}
while for $\alpha\approx -i\beta$
\begin{align}
    \text{Im}(k_{x}) \approx (\frac{3}{2}t)^{-1} 2\Delta^{2}((k_{y}-\mu)\mu)^{-\frac{1}{2}},
\end{align}
Thus, the Green's function has two components which have different decay lengths. However, due to the prefactor $(\alpha+i\beta)^{-1}$, the first contribution is highly suppressed compared to the second one, by a factor $\frac{\Delta^{2}}{\mu(p_{y}-\mu)}$. Therefore, the localization is determined by the second term,
\begin{align}
    \langle x \rangle \approx \frac{3}{2}t((k_{y}-\mu)\mu)^{\frac{1}{2}}\Delta^{-2}.
\end{align}
Thus, in this limit the localization length increases with increasing $k_{y}$ and is much larger than a superconducting coherence length, indicating that these states are similar to bulk states.

The localization can also be computed numerically from the results of our tight-binding model using a model with 1024 sites. Thus, the integration should be over the 512 sites on the left (right) if the state is more localized to the left (right). The results are shown in the main text, Fig. 3(b). The two lowest energy states have the same localization. The numerical results confirm the analytical calculations above, the localization length decreases as $k_{y}$ approaches $k_{F}$ and then strongly increases, approaching the bulk value of 256. For the two lowest energy states the localization is well determined. For the two next lowest energy states there is a clear oscillation for $k_{y}\gg k_{F}$. The reason for this is is that they become so close to the conduction band that they mix with those states.
\subsection{Finite bandwidth}\label{sec:FiniteBandwidth}
In previous sections, we made the assumption that $\mu/t\gg 1$ and used only the lowest order approximation. Here we  discuss how a finite value of $\mu/t$ influences the results by going to next order. 
We consider the correction to energy of the edge states to lowest order in $\mu/t$. To do this, we need to use higher order terms in $f(\vec{k})$ and $s(\vec{k})$ in their expansion around the $\pm\vec{K}$-points.
Taking into account next order terms in $f(k)$ and $s(k)$, they read
\begin{align}
    f(k) =  p_{x}+ip_{y}+\frac{3}{4t}(p_{x}^{2}-p_{y}^{2}+ ip_{x}p_{y}),\\
    \Tilde{f}(k) = - p_{x}+ip_{y}+\frac{3}{4t}(p_{x}^{2}-p_{y}^{2}- ip_{x}p_{y}),\\
    s(k) = \frac{3}{2}\sqrt{3}+\frac{9}{4}\sqrt{3}(p_{x}^{2}-p_{y}^{2}).
\end{align}
We will exploit the symmetry of the poles. Terms that are either odd in valley, or have opposite sign for the two poles within a single valley cancel out.

The corrections to the Green's function can be divided into two categories based on the decomposition introduced in Eq. B36. One is due to the change in the matrices $AG_{1}+BG_{2}$ introduced in Eq. B36 at each pole. The other one due to the change evaluation of the residue of $1/X$ in Eq. B36.
In the following we consider the corrections to each element. 

To consider the change in the residue of $1/X$ in Eq. B36 we must consider the correction to the location of the poles. The correction $\gamma$ to $k_{x}$ can be calculated by evaluating the zeros of $f$ taking into account the second order terms. To lowest order this correction is
\begin{align}
    \gamma = \mp\frac{3}{4t}((\alpha+i\beta)^{2}-k_{y}^{2}).
\end{align}
 For estimating the correction to the residue of $1/X$ at the poles we may use that
\begin{align}
    \lim_{p\xrightarrow{}p_{i}}\frac{p-p_{i}}{X}= \left(\frac{dX}{dp}|_{p = p_{i}}\right)^{-1}.
\end{align}
Now,
\begin{align}
    \frac{dX}{dp_{x}} &= (A-2\mu^{2})\frac{d}{dp_{x}}(F^{2})+A\frac{\Delta_{0}^{2}}{t^{2}}\frac{d}{dp}s^{2}
    \approx (A-2\mu^{2})\partial_{p_{x}}(k_{x}^{2}+k_{y}^{2}+(\frac{3}{2}p_{x}p_{y}^{2}-\frac{3}{4}p_{x}^{3}))\nonumber\\
    &=(A-2\mu^{2})(p_{x}+\frac{3}{2t}k_{y}^{2}-\frac{9}{4t}k_{x}^{2})\approx(A-2\mu^{2})(\alpha+i\beta\mp\frac{3}{2t}(\alpha+i\beta)^{2}),
\end{align}
where the derivative of $s(\vec{K})$ has been omitted because it is of higher order in $(\Delta/t)$.
With this, the residue of $1/X$ in the K-valley is
\begin{align}
    \frac{1}{\alpha\beta}\frac{\alpha\mp i\beta}{\alpha^{2}+\beta^{2}(1-\frac{3}{2t}(\alpha+i\beta))}\approx \Bigg(\frac{\alpha\mp i\beta}{\alpha^{2}+\beta^{2}}\pm\frac{3}{2t}\Bigg)\frac{1}{\alpha\beta},
\end{align}
that is, the correction equals $\pm\frac{3}{2t}\frac{1}{\alpha\beta}$.
In the opposite valley, the corrections differ by a minus-sign, making the correction even in $\vec{k}$. 
 Now we may exploit the symmetry of the poles. First, we may use that terms for which the corrections to the residue of $X^{-1}(AG_{1}+BG_{2})$ have opposite sign cancel out. 
 Because the corrections to the residue of $1/X$ have opposite sign within a single valley, only terms proportional to $k_{x}$ survive. This implies that the correction is proportional to $\rho_{x}$.
 Moreover, since $s(\vec{k})$ is odd in $\vec{k}$, the Nambu off-diagonal $\rho_x$- components are not corrected up to first order in $\frac{\mu}{t}$. Thus, the only correction is the $\rho_{x}$-contribution to the diagonal elements in Nambu space. It equals

\begin{align}
    &-\frac{3\mu}{2t\alpha\beta}\Bigg((\omega-i\sqrt{\Delta^{2}-\omega^{2}})(\alpha+i\beta)-(\omega+i\sqrt{\Delta^{2}-\omega^{2}})(-\alpha+i\beta )\Bigg)\rho_{x}\nonumber\\
    &=-\frac{3\mu}{t\alpha\beta}\Bigg(\alpha\omega+\beta\sqrt{\Delta^{2}-\omega^{2}}\Bigg)\rho_{x}.
\end{align}

Now consider the correction to $AG_{1}+BG_{2}$ in Eq. B36. For this correction we need to take into account the correction to $f$,
\begin{align}
    f\approx \alpha+i\beta+\frac{3}{2t}\Bigg((\alpha+i\beta)^{2}-p_{y}^{2}\Bigg)+\frac{3}{4t}(\alpha+i\beta)p_{y},\\
    \Tilde{f}\approx -\alpha+i\beta+\frac{3}{2t}\Bigg((\alpha+i\beta)^{2}-p_{y}^{2}\Bigg)-\frac{3}{4t}(\alpha+i\beta)p_{y}.
\end{align}
According to the same symmetries as before, only the $\rho_{x}$-component diagonal in Nambu space may survive, and it equals
\begin{align}
    &\frac{3\mu}{2t\alpha\beta(\alpha^{2}+\beta^{2})}\Bigg((\alpha-i\beta)(\omega-i\sqrt{\Delta^{2}-\omega^{2}})((\alpha+i\beta)^{2}-k_{y}^{2})+(\alpha+i\beta)(\omega+i\sqrt{\Delta^{2}-\omega^{2}})((\alpha-i\beta)^{2}-k_{y}^{2})\Bigg)\nonumber\\
    &=\frac{3\mu}{t\alpha\beta}\Bigg(\alpha\omega+\beta\sqrt{\Delta^{2}-\omega^{2}}\Bigg)+\frac{3\mu}{2t\alpha\beta(\alpha^{2}+\beta^{2})}\Bigg(\alpha\omega-\beta\sqrt{\Delta^{2}-\omega^{2}}\Bigg)k_{y}^{2}.
\end{align}
Summing the two corrections, we see that two of the three terms cancel against each other. Hence the only correction of first order in $\frac{\mu}{t}$ is
\begin{align}
    \frac{3\mu}{2t\alpha\beta(\alpha^{2}+\beta^{2})}\Bigg(\alpha\omega-\beta\sqrt{\Delta^{2}-\omega^{2}}\Bigg)k_{y}^{2}\rho_{x}.
\end{align}
That is, for $k_{y} = 0$ the correction vanishes and there is still a zero energy bound state.
Moreover, from this expression it follows we need to evaluate the $\beta$-term to have any nonzero energy bound states, which means that there exist no corrections to the energy of order $\mu/t$, lowest order corrections are of order $\Delta/\mu \mu/t = \frac{\Delta}{t}$.
Lastly, the correction to the energy is second order in $k_{y}$, and thus for $k_{y}\ll k_{F}$ this corresponds to a  negligible correction.
\end{document}